\documentclass[aps,pre,preprint, superscriptaddress,nofootinbib]{revtex4-1}
\usepackage{graphicx}
\usepackage{amsmath}
\usepackage{amsfonts}
\usepackage{hyperref}
\begin{document}
\title{Spatial structure of states of self stress in jammed systems}

\author{Daniel M. Sussman}
\affiliation{Department of Physics and Astronomy, University of Pennsylvania, Philadelphia, Pennsylvania 19104, USA}
\email[]{dsussman@sas.upenn.edu}
\author{Carl P. Goodrich}
\affiliation{School of Engineering and Applied Sciences, Harvard University, Cambridge, MA 02138, USA}
\author{Andrea J. Liu}
\affiliation{Department of Physics and Astronomy, University of Pennsylvania, Philadelphia, Pennsylvania 19104, USA}

\date{\today}

\begin{abstract}
States of self stress, organizations of internal forces in many-body systems that are in equilibrium with an absence of external forces, can be thought of as the constitutive building blocks of the elastic response of a material. In overconstrained disordered packings they have a natural mathematical correspondence with the zero-energy vibrational modes in underconstrained systems. While substantial attention in the literature has been paid to diverging length scales associated with zero- and finite-energy vibrational modes in jammed systems, less is known about the spatial structure of the states of self stress. In this work we define a natural way in which a unique state of self stress can be associated with each bond in a disordered spring network derived from a jammed packing, and then investigate the spatial structure of these bond-localized states of self stress. This allows for an understanding of how the elastic properties of a system would change upon changing the strength or even existence of any bond in the system.
\end{abstract}

\maketitle
\section{Introduction}
Jammed sphere packings -- collections of athermal particles interacting only via repulsive nearest-neighbor interactions -- are a simple and well-studied model of disordered solids \cite{Ohern2003}. The jamming transition, the packing fraction at which the system acquires nonzero elastic moduli, is isostatic with precisely the number of interparticle interactions needed to constrain all of the degrees of freedom in the thermodynamic limit. The distance of a given system from isostaticity can be characterized by the number of interacting neighbors per particle relative to the isostatic value, $\Delta z$. Much effort has gone into studying the diverging length scales that appear near the jamming critical point, e.g., those that control system stability with respect to free \cite{Wyart2005, Goodrich2013} and generalized boundary conditions \cite{Schoenholz2013, Sussman2015}; those that characterize the length scale over which system response to imposed forces is mediated \cite{Ellenbroek2006, Wyart2010, Lerner2014}; and those that characterize the propagation of nonlinear shocks \cite{Upadhyaya2014}. An emerging consensus is that above the transition there are at least two separately diverging length scales upon the approach to the critical point, which scale as $l_T \sim \Delta z^{-1/2}$ and $l^* \sim \Delta z^{-1}$, respectively, where $\Delta z$ vanishes at the jamming transition.

An understanding of these length scales is crucially important to the ability to design novel metamaterials, which rely on the ability to (locally) tune the elastic response of disordered systems by carefully engineering or modifying particle positions or bond constraints. For instance, one may want to selectively remove bonds to target a particular global elastic response \cite{Goodrich2015}, control where a structure will focus stress and buckle \cite{Paulose2015}, or apply boundary constraints to create highly nonlinear couplings between different global deformations \cite{Florijn2014}. While length scales related to vibrational modes have been well-characterized, much less attention has been paid to correlations of stress among the bonds of these disordered networks; clearly, though, such stress-stress correlations are equally important in determining and tuning the elastic properties of these systems. Here we will focus on ``states of self stress'' (SSS):  arrangements of stresses on the bonds that lead to zero net force on every particle.

In jammed systems prepared, e.g., by a compression algorithm, there is a special linear combination of the system's states of self stress in which all of the bond tensions have the same sign; this special state corresponds to the actual stress in the system and is related to the qualitatively different behavior of the bulk and shear moduli near the transition~\cite{Wyart:2005vu}. In the language of force network ensembles the force chains in a system are linear combinations of the states of self stress of a packing, and the statistics of SSS could inform observations of cluster sizes in force chains \cite{Nienhuis2006,Kondic2015}. In this paper we characterize the spatial extent of states of self stress and find a surprising result: their size scales neither as $\Delta z^{-1/2}$ nor as $\Delta z^{-1}$, but is more consistent with a scaling of $\Delta z^{-0.8}$ in $d=2$ and $\Delta z^{-0.6}$ in $d=3$, where $d$ is the spatial dimension. 

We begin by considering a set of $N$ particles connected by $N_b$ bonds in $d$ dimensions. The Maxwell-Calladine count \cite{Calladine1978}, which relates rigidity to isostaticity by connecting the number of particles, the number of bonds, the number of SSS, $N_S$, and the number of zero modes (motions of particles that to linear order do not stretch any bond), $N_0$, states that 
\begin{eqnarray}
N_0 = dN -N_b+N_S.
\end{eqnarray}
A system without external constraints has at least $f(d)$ zero modes, where $f(d) = d(d+1)/2$ for a system with free boundary conditions and $f(d)=d$ under periodic boundary conditions \cite{Maxwell1865}. These trivial zero modes represent global translations (and rotations for free boundary conditions) and thus do not affect the rigidity of the system. In contrast, any additional zero modes would make at least part of the system non-rigid. Thus the isostatic point, where $N_0$ and $N_S$ are at their minimum values of respectively $f(d)$ and $0$, can only be obtained when the number of bonds is $N_{b,c} = dN-f(d)$. This is the minimum number of bonds needed to render the system rigid to linear order. Defining the average number of contacts per particle as $z\equiv 2N_b/N$, the jamming critical point is at $z_c = 2d-2f(d)/N$, and the distance to isostaticity is typically written $\Delta z = z-z_c$. Note that systems that are over-constrained (i.e. above isostaticity) will necessarily possess SSS. 

In this work we focus on bead-spring networks in the absence of any pre-stresses on the bonds, i.e. on the so-called unstressed version of the jammed networks \cite{Alexander1998, Silbert2005pre}. To understand the mathematical connection between zero modes and states of self stress, let $e$ be a vector of bond tensions/compressions, with $e_i$ denoting the stress in bond $i$. The equilibrium matrix, $Q$, relates this $N_b$-vector of bond stresses to the $dN$-vector of net force loads on the particles, $l$, via $Qe=l$ \cite{Pellegrino1993}. The compatibility matrix, $Q^T$, relates the $dN$-vector of particle displacements, $x$, to the $N_b$ vector of bond strains, $s$, via $Q^T x=s$. The flexibility matrix $F$ connects the stresses to the strains, $s= F e$, and for the spring networks considered in this work $F$ is a diagonal matrix of spring constants, $F_{ii} = 1/k_i$. With this notation the energetic cost of a set of particle displacements can be written as 
\begin{eqnarray}
E &=&\frac{1}{2}s^T e \nonumber \\
&=&\frac{1}{2}x^T Q F^{-1} Q^T x, 
\label{eq:energetic_cost_of_displacement}
\end{eqnarray}
and we see that the matrix $\mathcal{M} = QF^{-1}Q^T$ is the dynamical matrix of the unstressed system. 
Similarly, the energetic cost of a set of bond strains is given by
\begin{eqnarray}
E &=&\frac{1}{2} s^T F^{-1} s \nonumber \\
&=& \frac{1}{2} e^T F e.
\label{eq:energetic_cost_of_strain}
\end{eqnarray}
In the following we will take all of the springs to have unit stiffness, $k_i = 1$. 

A zero mode is defined as a set of particle displacements that does not strain any bonds. Since the strain is $s^T = xQ$, we see that the zero modes are the left singular vectors of $Q$, or equivalently they are elements in the null space of $\mathcal{M}$. Similarly, a state of self stress is defined as a set of bond stresses that does not create any net forces on the particles. Since $l=Qe$, we see that the states of self stress are the right singular vectors of $Q$, or equivalently elements of the null space of the $N_b\times N_b$ matrix $\mathcal{N}=Q^T Q$. The matrix $\mathcal{N}$, and in particular its positive-frequency spectrum, has previously arisen as a ``rheology operator'' in the context of non-Brownian suspension flows \cite{Wyart2012}; here we are interested only in the SSS and hence the null space of the operator. The operator has also appeared as the supersymmetric partner to the dynamical matrix in recent theories of topologically protected phonons in ordered and disordered systems \cite{LubenskyRev, Lubensky2014, SussmanTopo}.

That the null space of $\mathcal{N}$ is composed of the states of self stress can also be understood by the following physical argument. We imagine that all of the particles are pinned to their initial position by fictitious springs (of strength $\tilde{k}_i=1$). Using the connection between applied stresses and particle displacements above, a set of external stresses applied to the system's bonds could result in particle displacements, stretching these fictitious springs. The energetic cost of the fictitious springs can be written as
\begin{equation}
E_b = \frac{1}{2} e^T Q^T Q e.
\end{equation}
Note that this is \emph{not} the total energetic cost imposed by the stresses in general: if the stresses cause a change of the position of the particles there would be an additional contribution from the stretching of the original bonds of the system. However, this makes it clear that the null space of $\mathcal{N} = Q^T Q$ gives the states of self stress: if $E_b=0$ then the fictitious springs have not stretched and hence no particle has been displaced from its initial position by the imposed bond stresses. Thus, $e$ imposed no net load on any particle -- precisely the definition of a SSS.

The states of self stress of a system form a convenient basis to understand the energetic cost of imposed deformations. To see this, note that a global deformation, such as compression or shear, can be represented by an imposed strain on each bond. If this strain projects entirely onto the states of self stress, then force balance will be satisfied and the energy of the deformation is given by Eq.~\eqref{eq:energetic_cost_of_strain}. Generically, however, the imposed strain does not entirely project onto the SSS, leading to a net load on the particles. This results in a secondary response of the particles to regain force balance that precisely relaxes the part of the stress that does not project onto the SSS. Therefore, as discussed in more detail in Ref. \cite{LubenskyRev}, the elastic energy of such a deformation can be expressed completely in terms of the projection of the imposed strain onto the SSS of the system.

With the operator $Q$ and its kernel in hand, we will proceed to investigate the localization of SSS associated with individual bonds of jammed packings. Just as removing a bond from a sub-isostatic network produces a new zero mode whose spatial extent depends critically on $\Delta z$, the Maxwell-Calladine count tells us that adding a bond to a hyperstatic network will generically lead to a new state of self stress (here we assume that a hyperstatic lattice with no rattlers has exactly $f(d)$ zero modes and a hypostatic lattice has no states of self stress). Equivalently, the removal of a bond from such a network will in general reduce the dimension of $\textrm{ker}(\mathcal{N})$ by one, and the element that vanishes from the null space identifies a SSS that depends on the existence of that particular bond. 

In this paper we set up a framework for thinking about the spatial organization of states of self stress in disordered networks. Specifically, we will see that there is a unique state associated with each bond defined by the component of the kernel of $Q$ that is lost when the bond is removed. This state allows one to calculate directly what would happen to the elasticity of the system should the bond be removed~\cite{LubenskyRev}, which is a key step in designing materials with tunable elastic properties~\cite{Goodrich2015}. The localization of these states also plays a role in the width of the fracture zone of marginal materials, as observed by Driscoll {\it et al.}~\cite{Driscoll:2015vf}. Very recently the idea of the state of self stress associated with a bond added to precisely isostatic system was used to construct a new variational argument to explain the density of vibrational modes near the jamming transition \cite{Wyart2016}. We anticipate that the details of the spatial structure that we find for the states of self stress may directly influence this argument.

The remainder of the paper is organized as follows. Section \ref{sec:methods} gives further details on the systems studied and the techniques used to study the SSS. Section \ref{sec:spatial} shows the results of our investigations in two and three dimensions, and we close in Section \ref{sec:disc} with a discussion of our results.

\section{Model and Methods}\label{sec:methods}
In this work we focus on numerically generated jammed packings of $N$ spheres in two and three dimensions. The systems are all polydisperse mixtures with a flat distribution of particles sizes between $\sigma$ and $1.4\sigma$, where $\sigma$ represents the smallest particle diameter. The particles interact via a finite-ranged soft harmonic potential, 
\begin{equation}
V(r_{ij})=\left\{ \begin{array}{cr} \frac{\epsilon}{2}\left( 1-r_{ij}/\sigma_{ij}  \right)^2\quad & r_{ij}<\sigma_{ij} \\ 0 & r_{ij}\geq\sigma_{ij} \end{array} \right. ,
\end{equation}
where $r_{ij}$ is the distance between particle centers, $\sigma_{ij}$ is the sum of their radii, and $\epsilon$ sets the energy scale. In preparing the systems we take all particles to have equal mass $m$, and since we are interested in states of stress in which there are no net forces on the particles we expect this to have no effect on the results reported below. Throughout the paper we will measure distance in units of the average particle diameter, $\langle \sigma \rangle$, energy in units of $\epsilon$, and pressure in units of $\epsilon/\langle \sigma \rangle^{d-1}$.

Our disordered configurations were obtained by preparing jammed states at a target pressure, $p$: particles were initially placed at random in the simulation box with linear dimensions $L$ and with periodic boundary conditions (i.e. in an infinite temperature configuration), and then quenched to zero temperature by combining linesearch methods, Newton's method, and the FIRE algorithm \cite{quench}. The systems were then incrementally expanded or compressed uniformly and then re-quenched to zero temperature until the target pressure was obtained to within $1\%$.  The average value of $\Delta z$ is directly related to the pressure at which the jammed systems were initially prepared; for these harmonic interactions $\Delta z \sim p^{1/2}$ \cite{Ohern2003}. The bulk of the work in this manuscript used systems of size $N=8192$ in two dimensions and $N=12000$ in three dimensions.

In detail, our procedure for identifying SSS associated with particular bonds is as follows. We begin with a jammed packing (with periodic boundary conditions) and remove all ``rattlers'' (particles with less than $d+1$ contacts) -- for jamming-derived networks all remaining bonds will participate in at least one SSS, and so this ensures that removing a bond from the network reduces $\textrm{dim}\left( \textrm{ker} (\mathcal{N})\right)$ by one rather than introducing a new zero mode. From this system we construct the operators $Q$ and $\mathcal{N}$ and numerically obtain the null space of $\mathcal{N}$ using the ARPACK package \cite{arpack}. Despite ARPACK's ability to efficiently find extremal eigenvectors, and although we only were interested in obtaining the null space of $\mathcal{N}$, we consistently found that for our large, high-pressure systems (with their very large, degenerate null space) we needed to perform a complete diagonalization of $\mathcal{N}$ in order to obtain a numerically accurate basis of states of self stress. This remains the greatest impediment to studying much larger systems by this method.

With an orthonormal basis for the SSS, we  choose a bond, $b_i$, and any element in the null space of $\mathcal{N}$, $e$, with a non-zero projection onto $b_i$. We subtract off enough of $e$ from every other element of the null space so that only $e$ has a non-zero component $e_i$. Finally, we implement a standard modified Gram-Schmidt method to orthonormalize $e$ and the remainder of the null space of $\mathcal{N}$. After this procedure we have a unique single state of self stress, $e$, that would vanish from the kernel if bond $b_i$ were removed, whereas all other elements of the null space would be unaffected. To within numerical precision the obtained state of self stress is independent of the randomly chosen initial state. We note that we can perform this procedure for every bond in the system, even though the number of bonds in the system is larger than $\textrm{dim}\left( \textrm{ker} (\mathcal{N})\right)$. This immediately implies that the unique state of self stress associated with bond $i$ will not, in general, be orthogonal to the unique SSS associated with bond $j$. In the appendix we explore an alternate definition of the state of self stress associated with a particular bond, based on exploring the landscape of linear combinations of $e$ with other elements of  $\textrm{dim}\left( \textrm{ker} (\mathcal{N})\right)$.

\begin{figure}
\centerline{
\includegraphics[width=.4\linewidth]{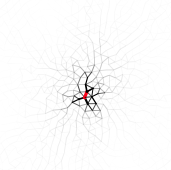}
\includegraphics[width=.4\linewidth]{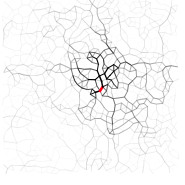}
}
\caption{\label{fig:stressfig} Typical states of self stress in a 2D $N=1024$ system prepared at $\Delta z=1.97$ (left) and  $\Delta z=0.197$ (right). Shown is the state of self stress that uniquely requires the presence of the thickest bond (labeled in red). Bond thickness in these images are proportional to the magnitude of the stress in the state.}
\end{figure}

Representative examples of states of self stress are shown in Fig. \ref{fig:stressfig}, where the selected bond is shown in red, and the magnitude of $e$ along each contact is illustrated by the thickness of the bond. Visual inspection of this figure suggests that the degree of localization of $e$ strongly varies with $\Delta z$, suggesting a growing length scale as the isostatic point is approached. Any localization of these SSS, i.e., any exponential decay of their average magnitude as a function of distance from the selected bond, presumably  crosses over to a power-law behavior at large distances. This is expected both on the grounds that the SSS contribute to the mechanical rigidity and thus presumably have the power-law tails expected from continuum elasticity, and also based on previous analyses that $l_T \sim \Delta z^{-1/2}$ controls the length scale at which continuum elastic properties set in for these harmonic jammed systems \cite{Lerner2014,During2013}.

\begin{figure}
\centerline{\includegraphics[width=.8\linewidth]{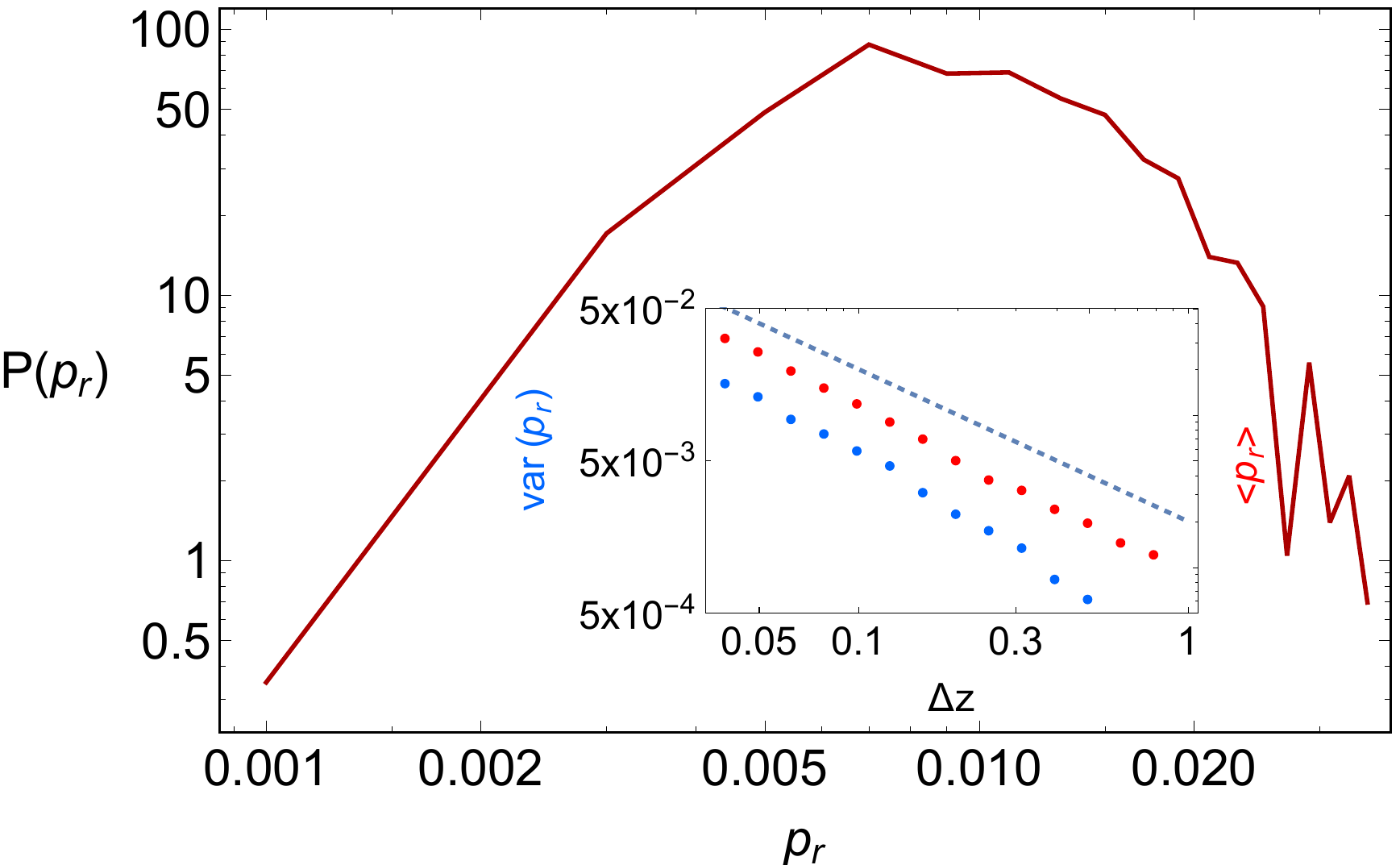}}
\caption{\label{fig:prplot} Distribution of participation ratios for the states of self stress $e$ for an $N=8192$ system in two dimensions prepared at $\Delta z = 0.099$. Inset. The mean (upper, red points) and standard deviation (lower, blue points) of $P(p_r)$ as a function of pressure. The dotted line is a guide to the eye with slope $-1/2$.}
\end{figure}

A straightforward if indirect measure of how extended these modes are comes from the participation ratio of the SSS $e$:,
\begin{equation}
p_r\equiv \frac{\left( \sum_i^{N_b} |e_i|^2 \right)^2}{N_b \sum_i^{N_b} |e_i|^4}.
\end{equation}
The participation ratio quantifies the degree to which the $e$ has dominant contributions from just a few bonds as opposed to being more uniformly extended over many bonds: $p_r = 1/N_b$ would corresponds to a SSS (unphysically) completely localized to a single bond, and $p_r=1$ would correspond to a SSS uniformly extended over all bonds in the system. Figure \ref{fig:prplot} shows the distribution of participation ratios obtained by computing the SSS $e$ associated with every bond of a single jammed packing with $N=8192$ in two dimensions. Both the mean and the standard deviation of this distribution has a scaling consistent with $\Delta z^{-1}$. This scaling, identical to the scaling of $l^*$, suggests the importance of that length scale in the spatial organization of the states of self stress. However, as we will see below, this measure is overly sensitive to the behavior of the long-range tail in the state of self stress, and so does not reflect the size of the localized region.

\begin{figure}
\centerline{\includegraphics[width=.8\linewidth]{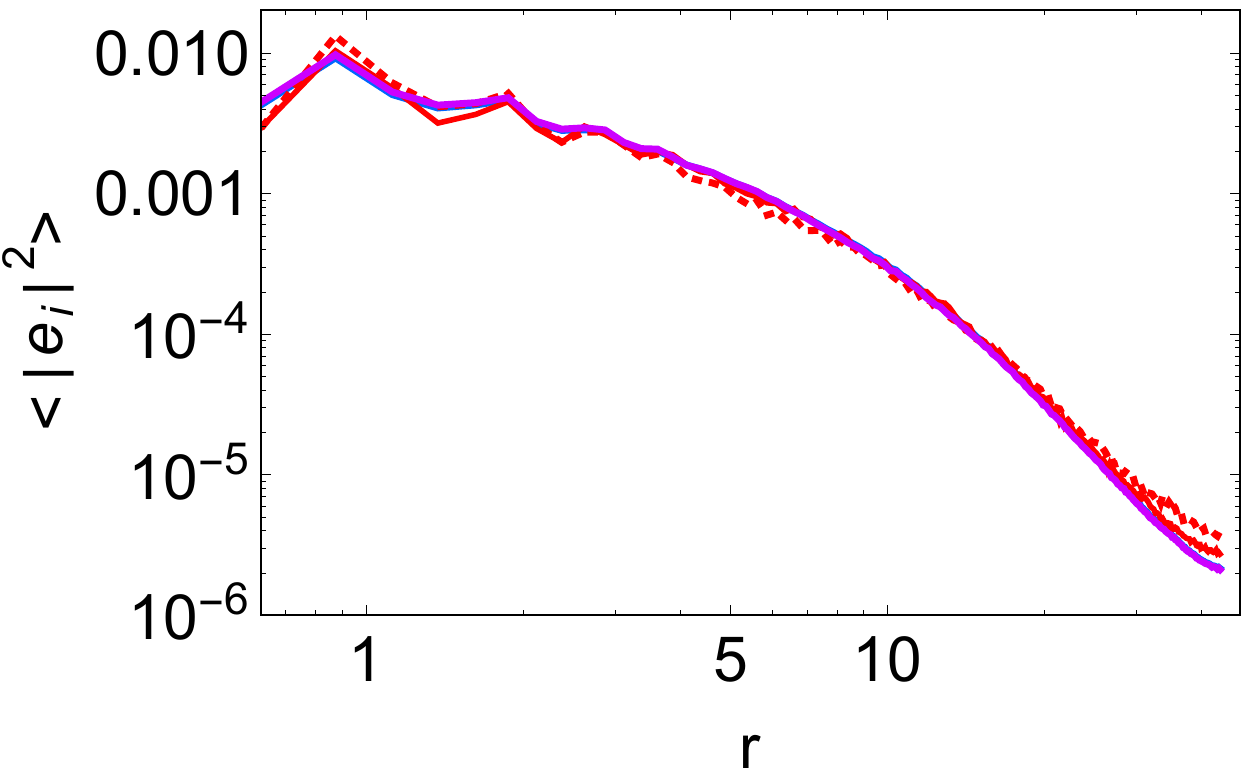}}
\caption{\label{fig:explot2d} Average radial profiles of states of self stress unique to bonds in a 2D system with $N=8192$ prepared at $\Delta z = 0.31$. Blue lines correspond to averages over all bonds in the system, purple lines correspond to averages over those bonds whose state of self stress has a low participation ratio (lower 80th percentile), solid red lines correspond to averages over bonds with a large contribution to the bulk modulus (upper 5th percentile), and dotted red lines correspond to averages over bonds with a large contribution to the shear modulus (upper 5th percentile). }
\end{figure}

Following Goodrich et al. \cite{Goodrich2015}, one can decompose the global bulk and shear modulus, $B$ and $G$, into contributions from individual bonds, e.g., $B_i$, such that $B=\sum_i B_i$ and $G=\sum_i G_i$. In general we find only very small correlations between $B_i$ and $p_r$ and between $G_i$ and $p_r$, correlations which additionally depend on the size of the system relative to the diverging length scales near the jamming transition. Representative examples are shown in Fig. \ref{fig:explot2d}, which plots the stress profiles of $e$ as a function of distance from the selected bond averaged over different subsets of bonds in the system.

\section{Spatial organization of states of self stress}\label{sec:spatial}
We now study in greater detail the average spatial profiles of the SSS as a function of dimension and distance to isostaticity of the jammed states we studied. The question of how exponentially localized these states are close to the targeted bond is complicated by the presence of many contributing length scales: at a minimum, $l^*\sim \Delta z^{-1}$ and $l_T\sim \Delta z^{-1/2}$ \cite{Wyart2005,Silbert2005} are likely to be present in the spatial profile of the modes \cite{Sussman2015}, and there is the additional complication that over very short length scales and independent of the pressure, out to $\sim 2 - 3\sigma$, the jammed packings have local structure, i.e., a radial distribution function $g(r)$ that is not flat. Particularly in three dimensions, then, it is difficult to both prepare and diagonalize systems large enough and over a wide enough range of pressure to explore the spatial profile over distances $r$ both large and small compared to all three of these lengths, as the computational complexity to compute $e$ grows like $\mathcal{O}(N_s^2N_b^2)$, and at fixed $\Delta z$ both $N_s$ and $N_b$ scale linearly with $N$. 

\begin{figure}
\centerline{\includegraphics[width=.75\linewidth]{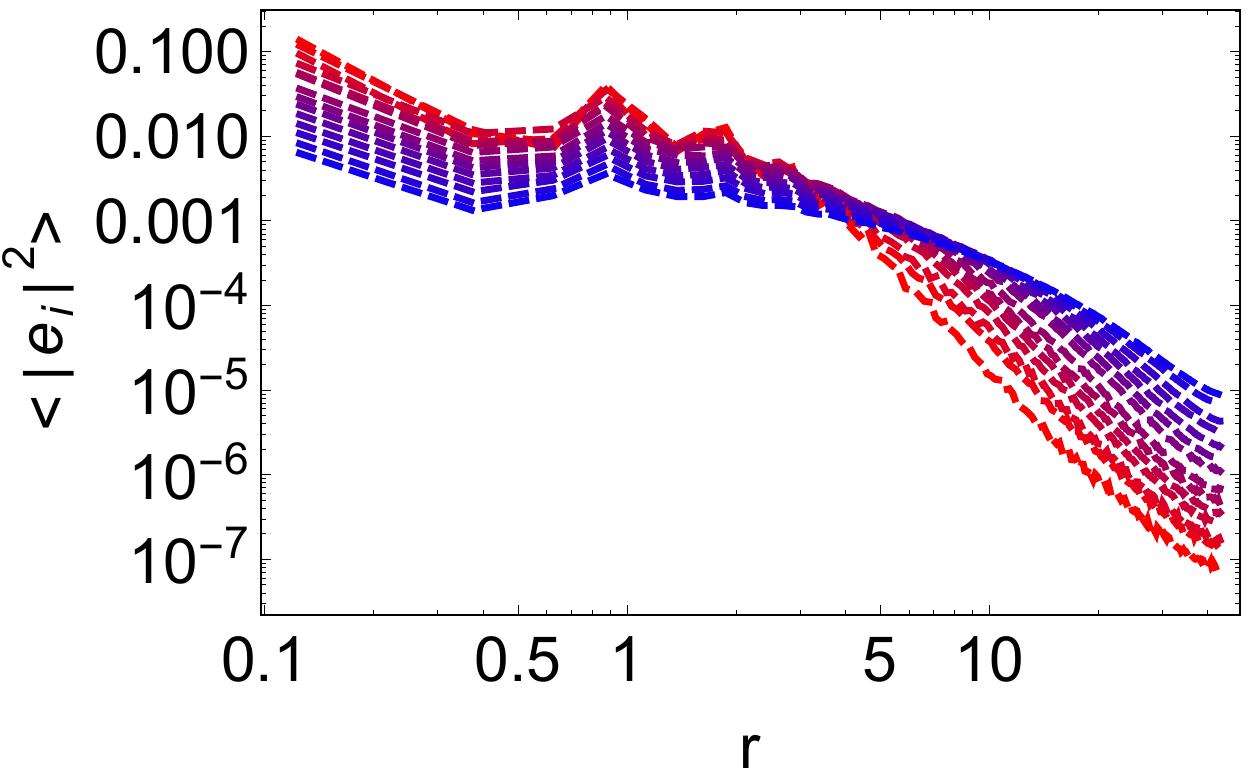}}
\caption{\label{fig:plot2d8192} Average radial stress profiles for 2D systems with $N=8192$ prepared at coordinations $\Delta z=0.039 - 0.78$ (increasing $\Delta z$ monotonically corresponds to increasing localization).}
\end{figure}

In two dimensions, the results of computing the $e$'s associated with the bonds of an $N=8192$ system and plotting the average spatial profile as a function of distance to the selected bond are shown in Fig. \ref{fig:plot2d8192}. The oscillations at small $r$ are indicative of the local structure of the packing over small distances; at larger $r$ there is a clear crossover in the decay, and this crossover moves to larger distances as the network approaches the critical point. 

\begin{figure}
\centerline{\includegraphics[width=.75\linewidth]{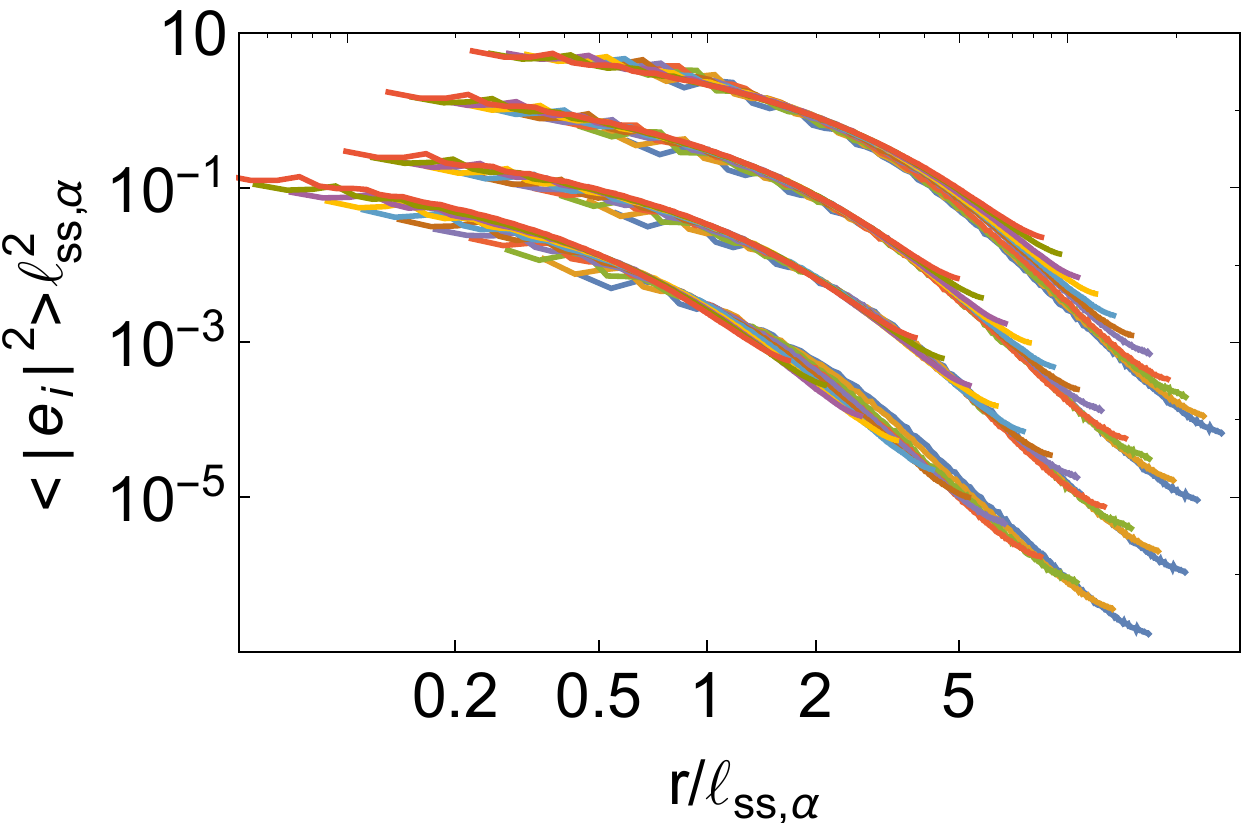}}
\caption{\label{fig:collapse2d} Average radial stress profiles for systems in $d=2$ with $N=8192$ prepared at $\Delta z=0.039 - 0.99$.  In each set of curves, the radial distance is scaled by a length $\ell_{ss,\alpha}=\Delta z^{-\alpha}$ while the y-axis is scaled by the volume $\ell_{ss,\alpha}^d$.  From top to bottom $\alpha = 1/2,\ 2/3,\ 3/4,\ 1$. Each set of curves has been vertically shifted for clarity.}
\end{figure}

This crossover is made more strikingly clear in Fig. \ref{fig:collapse2d}, where we attempt to collapse the curves by scaling the $x$-axis by a length scale defined by $\ell_{ss,\alpha}=\Delta z^{-\alpha}$ and the $y$-axis by $\ell_{ss,\alpha}^d$ ({\it i.e.}, assuming that the $e$'s have an exponentially localized structure in $d$ dimensions, we scale by the $d$-dimensional volume of this structure). Assuming that the localization occurs on a scale $l_T\sim \Delta z^{-1/2}$ (corresponding to $\ell_{ss,\alpha}$ with $\alpha=1/2$) as in the localized zero modes that are present below isostaticity \cite{During2013}, we find that scaling distances by $l_T$ in the average spatial profiles of $e$ leads to reasonably good collapse of the data. At both large and small distances, where we do not expect the exponential localization to hold, the curves are not collapsed; since the vectors $e$ are normalized this leads to a modest vertical spreading of the curves in this representation. Of note, we find that scaling distances by $l^* \sim \Delta z^{-1}$, as suggested by the scaling of $\langle P(p_r) \rangle$, leads to a pronounced spreading of the radial stress profiles. Surprisingly, we find that using an intermediate value of $\alpha$ leads to a much cleaner collapse of these curves: using either $\alpha =2/3$ or $\alpha =3/4$ leads to quite clean collapse in the crossover region, perhaps reflecting a competition between $l_T$ and $l^*$ in the self stress organization. Certainly the figure suggests that the scaling of $l_T$ and $l^*$ represent lower and upper bounds on the correct scaling behavior.

\begin{figure}
\centerline{\includegraphics[width=.75\linewidth]{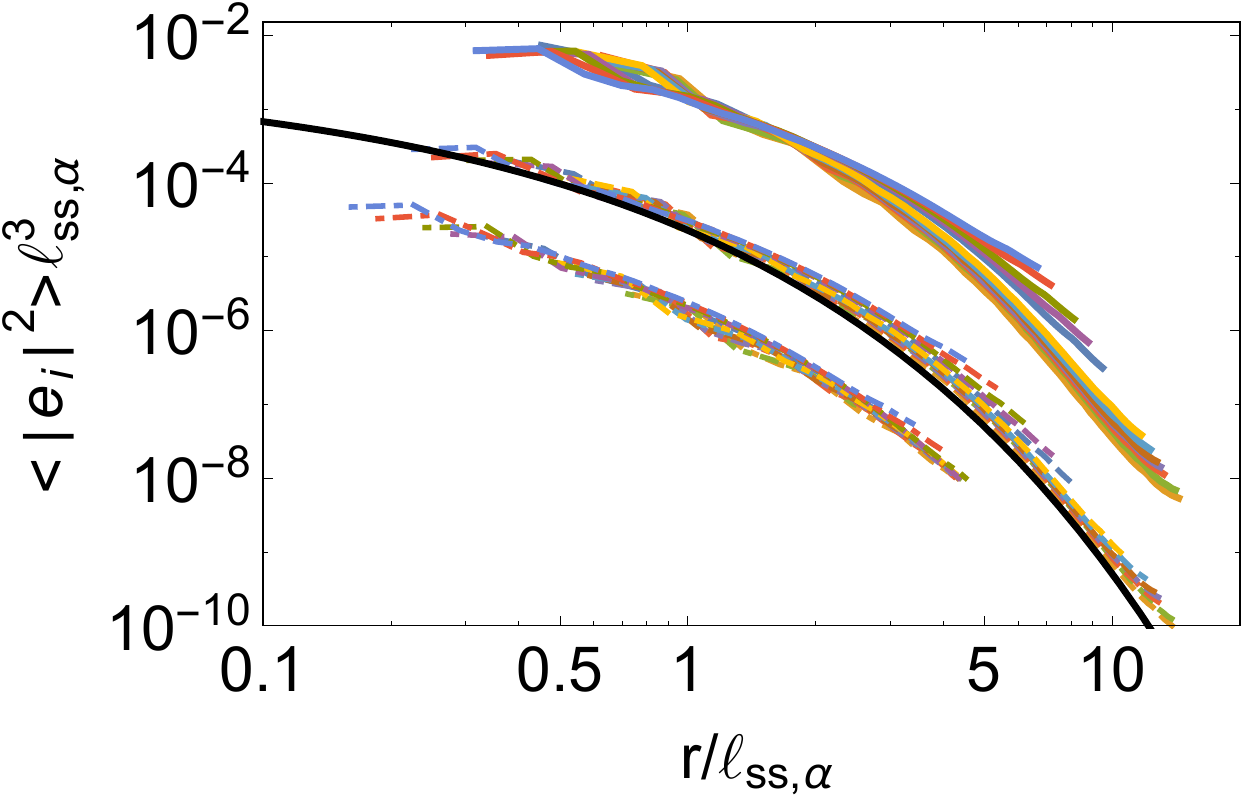}}
\caption{\label{fig:collapse3d} Average radial stress profiles for 3D systems with $N=12000$.  In each set of curves, the radial distance is scaled by a length $\ell_{ss,\alpha}=\Delta z^{-\alpha}$ while the y-axis is scaled by the volume $\ell_{ss,\alpha}^d$.  Upper curves correspond to $\alpha=1/3$, the middle curves (vertically shifted for clarity) correspond to $\alpha=1/2$, and the lower curves (vertically shifted for clarity) correspond to $\alpha=2/3$. The black line shows a stretched-exponential decay function fit to one of curves.}
\end{figure}

In three dimensions the physical picture is harder to extract: as noted above it is both slow and numerically difficult to study a similar range of linear system sizes -- $L$ grows more slowly with $N$ in three dimensions, and the computational cost grows with $N_b$ rather than with $N$ (i.e., the computational cost is exacerbated by the fact that the critical coordination number itself, $z_c =2d$, grows with dimension). Thus, to access systems spanning the range of $L$ being either large or small relative to $l_T$  we studied three-dimensional systems with $N=12000$  with $0.059 \leq \Delta z \leq 1.5$.  In Fig. \ref{fig:collapse3d} we attempt to collapse the curves of $\langle |e_i|^2 \rangle$ with $\ell_{ss,\alpha}$ for various values of $\alpha$. We find that the curves are reasonably well collapsed by either $\ell_{ss,\alpha=1/2}$ or $\ell_{ss,\alpha=2/3}$, although not quite as cleanly as in two dimensions.

\begin{figure}
\centerline{\includegraphics[width=.75\linewidth]{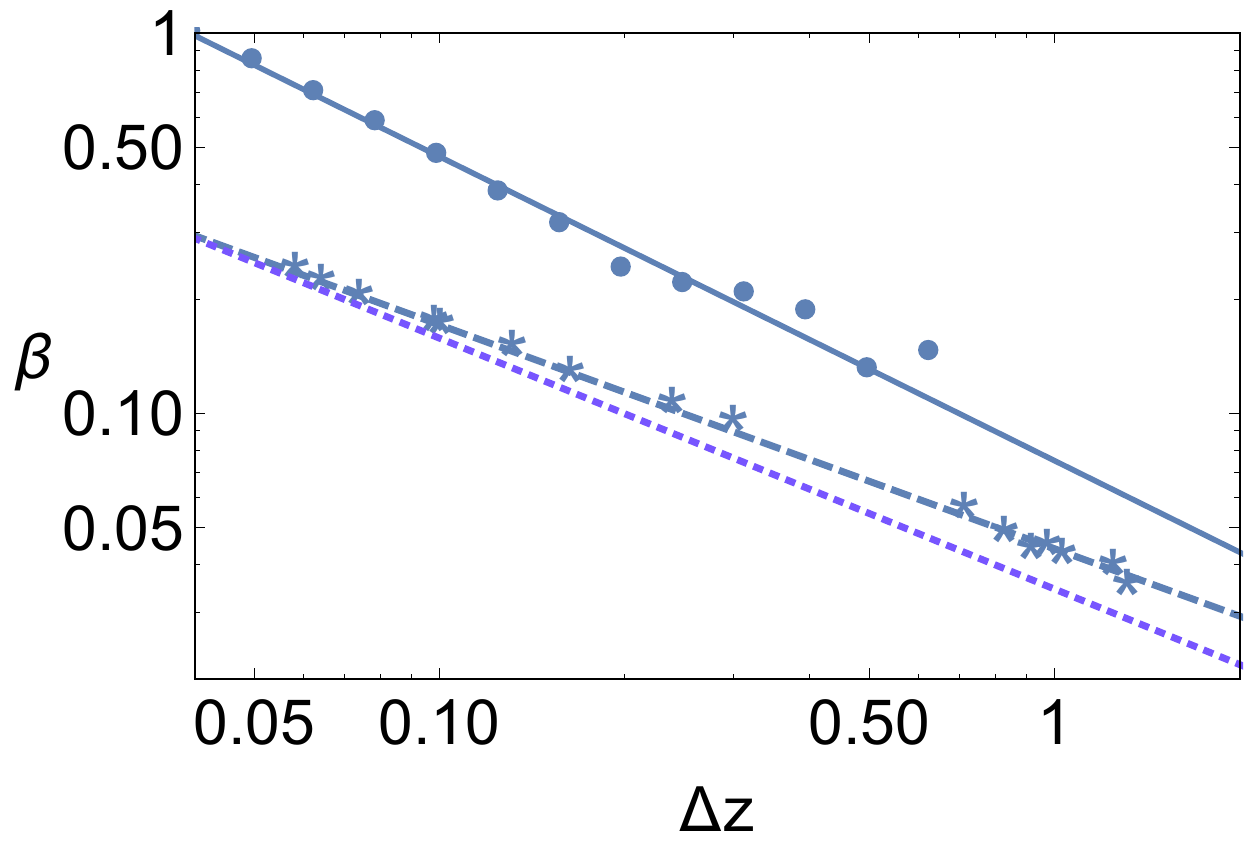}}
\caption{\label{fig:comblengths} Decay length as measured by fitting the radial stress profile of the $e$ to an exponential as a function of $\Delta z$. Blue dots and stars correspond to averages in 2D and 3D, respectively. Fit lines with slope $-0.8$ (upper, solid curve) and $-0.6$ (lower, dashed curve) are shown as guides to the eye. Also shown is a guide to the eye with slow $-2/3$ (dotted curve).}
\end{figure}

Clearly this data is system-size limited in trying to achieve good data collapse. As a second test, rather than scaling the average spatial profiles of the SSS by a $\Delta z$-dependent length scale, $\ell_{ss,\alpha}= \Delta z^{-\alpha}$, we attempt to fit the portion of $\langle |e_i|^2\rangle (r)$ after the regime dominated by local jammed structure (i.e., for $r> 2\sigma$), to a stretched exponential decay of the form $\langle |e_i|^2\rangle (r) = A \exp \left( -\left(r/\beta\right) ^\gamma \right)$. We repeat this for each of the pressures studied in both two and three dimensions, and a typical fit is shown as the solid black line in Fig. \ref{fig:collapse3d}. First, we note that choosing a stretching exponent of $\gamma = 1/2$ results in good fits for all dimensions and values of pressure. The values of $\beta$, the putative decay length scale, are shown in Fig. \ref{fig:comblengths}. (If the stretched-exponential form were a perfect fit, we would expect $\beta \sim \ell_{ss}$, where $\ell_{ss}$ is the length scale that best collapses the data for the radial profiles in Figs.~\ref{fig:collapse2d}-\ref{fig:collapse3d}.) We find that this data, too, is consistent with $\beta_{2d}\sim \Delta z^{-\alpha}$ with $\alpha \neq 1/2$ and $\alpha \neq 1$. Our best fit values suggest values closer to $\beta_{2d}\approx \Delta z^{-0.8}$ and $\beta_{3d}\approx \Delta z^{-0.6}$, but again the limited linear system sizes we are able to probe in three dimensions is an impediment to more accurately determining the exponent.

\section{Discussion}\label{sec:disc}
Our numerical results indicate the clear presence of a length scale over which states of self stress can be associated with individual bonds, $\ell_{ss}$. Surprisingly, the scaling of $\ell_{ss}$ with $\Delta z$ seems to be inconsistent with either of the most natural length scales near the jamming transition,  $l_T$ and $l^*$, but lies in between these two scalings. We have additionally presented data that suggests a dimension-dependent scaling of $\ell_{ss}$, although again we caution that in three dimensions the data is less clear. This is largely due to the much greater numerical difficultly of preparing very large systems and especially diagonalizing their relevant operators to find a complete basis for their states of self stress. While based on the scaling collapse in Figs. \ref{fig:collapse2d},\ref{fig:collapse3d} the simple relationship $\ell_{ss}\sim l_T$ cannot be completely ruled out, we believe our data is more consistent with a stronger scaling with $\Delta z$. The fact that a stretched exponential decay fits the profile of the $e$ above suggests that many length scales may be contributing, and one hypothesis for the scaling we observe is that the localization of the states of self stress is controlled by the geometric mean of the length scales $l_T$ and $l^*$, $\ell_{ss}\sim \sqrt{l_T l^*}$ or a dimensionality-weighted version, $\ell_{ss} \sim (l_T^{d-1} l^*)^{1/d}$.

Intriguingly, Fig. \ref{fig:comblengths} modestly hints at a slightly weaker dependence on $\Delta z$ in the three-dimensional data compared with the two-dimensional systems. This may be related to yet another dimension-dependent length scale associated with pressure fluctuations in jammed packings. Working in the force network ensemble \cite{Tighe2008, Tighe2010} a length scale $l_w \sim (l^*)^{1/d}$ was identified as the typical length over which pressure fluctuations in a stressed jammed packing persisted  \cite{Tighe2011}. This length can also be rationalized as a ``density of states of self stress'' in a system, and in this guise it would make some sense to find traces of $l_w$ in the spatial distribution of the localized states of self stress studied in this work. Nevertheless, while suggestive, we do not believe that our numerical evidence is sufficient to distinguish the small difference between, e.g.,  scaling by  $\ell_{ss}\sim \sqrt{l_T l^*}\sim \Delta z^{-3/4}$ or $\ell_{ss}\sim \sqrt{l_w l^*}\sim \Delta z^{-2/3}$ in three dimensions.

If this is indeed the case, though, it sets up an interesting tension, where the length scale signaling the onset of continuum elastic behavior in the vibrational and force-dipole response of the system is independent of $d$ yet the pressure fluctuations and typical localization lengths for the states of self stress, which again combine to control the elastic properties of the system, \emph{do} have a dependence on dimensionality. Resolving these issues will be of crucial importance when it comes to tuning and sculpting SSS by modifying bonds in disordered packings as a way to create interesting metamaterials. We hope that this work leads to a systematic exploration of these ideas in large three-dimensional systems where any possible dimension-dependence of $\ell_{ss}$ may be elucidated. One approach to doing this would be to modify the method of D{\"u}ring et al. \cite{During2013}, which studied the localization of zero-energy vibrational modes in sub-isostatic systems by studying the response of overdamped floppy networks in response to force dipoles; the analogous method here would involve replacing the springs by dashpots and studying the response of the network to an instantaneous strain along $b_i$.

We note that here we have explored only one aspect of these bond-localized SSS, i.e., their average spatial profiles. Of particular interest would be to connect the spatial organization of individual states of self stress with the contribution of that bond to the bulk or shear moduli of the system. We have seen above that the participation ratio is not sufficiently sensitive to the localization length of the states of self stress, but for the system sizes studied the non-averaged spatial profiles of the $e$ are too noisy to usefully analyze. Pursuing alternate measures of the localization of the $e$ and studying the associated distributions and correlations with $B_i$ and $G_i$ would be a natural and interesting extension of this work.

We also anticipate that these different force networks could also be fruitfully analyzed in terms of recent developments in the topological characterization of disordered networks. One interesting approach would be to apply recent techniques from computational algebraic topology to further investigate the fundamental structure of the disordered stress patterns \cite{perhom}. These tools, however, are computationally expensive, and a separate  approach that has been applied to  granular systems is the use of ``geographic community structures'' from network sciences \cite{Bassett2012}. Additionally, the recent efforts to characterize networks in terms of their loopiness and hierarchical organization may shed light on the underlying structure of these states of self stress in both two \cite{Katifori2012} and three dimensions \cite{Katifori2015}.

Our work gives a general framework for thinking about the spatial organization of states of self stress in jammed systems that should be particularly relevant for understanding the effect of making changes to a network. For example, the unique state associated with each bond can be used to predict changes in the elastic constants upon bond pruning and can be used to design materials with specified elastic properties~\cite{Goodrich2015}. In addition, our results shed light on the work by Driscoll {\it et al.}~\cite{Driscoll:2015vf}, which studied fracture in disordered networks under uniaxial tension. By allowing bonds to break when the bond's stress exceeds a threshold, they find cracks with a failure zone of width $w$. Intuitively the size of the failure zone should be related to $\ell_{ss}$, because the unique state of self stress associated with a failing bond determines precisely the set of bonds among which the remaining stress must be redistributed. However, in simulations of square lattices with randomly added next-nearest neighbors Driscoll {\it et al.} find that $w \sim \Delta z^{-1/2} \sim l_T < \ell_{ss}$ for large fixed $L$~\cite{Driscoll:2015vf}. It is not clear if the small difference in observed exponents can be reconciled within the accuracy of the respective sets of data, if it is related to the difference between jammed systems and square lattices with random crossbars, or if the bond-breaking simulation is complicated by additional physics.

We thank Bryan Chen, Anton Souslov, and Sidney Nagel for useful discussions. This research was supported by the Advanced Materials Fellowship of the American Philosophical Society (DMS) and by the US Department of Energy, Office of Basic Energy Sciences, Division of Materials Sciences and Engineering under Award DE-FG02-05ER46199 (AJL, CPG). 
This work was partially supported by a Simons Investigator award from the Simons Foundation to AJL and a University of Pennsylvania SAS Dissertation Fellowship to CPG.

\bibliography{selfstress_bib}

\begin{thebibliography}{37}%
\makeatletter
\providecommand \@ifxundefined [1]{%
 \@ifx{#1\undefined}
}%
\providecommand \@ifnum [1]{%
 \ifnum #1\expandafter \@firstoftwo
 \else \expandafter \@secondoftwo
 \fi
}%
\providecommand \@ifx [1]{%
 \ifx #1\expandafter \@firstoftwo
 \else \expandafter \@secondoftwo
 \fi
}%
\providecommand \natexlab [1]{#1}%
\providecommand \enquote  [1]{``#1''}%
\providecommand \bibnamefont  [1]{#1}%
\providecommand \bibfnamefont [1]{#1}%
\providecommand \citenamefont [1]{#1}%
\providecommand \href@noop [0]{\@secondoftwo}%
\providecommand \href [0]{\begingroup \@sanitize@url \@href}%
\providecommand \@href[1]{\@@startlink{#1}\@@href}%
\providecommand \@@href[1]{\endgroup#1\@@endlink}%
\providecommand \@sanitize@url [0]{\catcode `\\12\catcode `\$12\catcode
  `\&12\catcode `\#12\catcode `\^12\catcode `\_12\catcode `\%12\relax}%
\providecommand \@@startlink[1]{}%
\providecommand \@@endlink[0]{}%
\providecommand \url  [0]{\begingroup\@sanitize@url \@url }%
\providecommand \@url [1]{\endgroup\@href {#1}{\urlprefix }}%
\providecommand \urlprefix  [0]{URL }%
\providecommand \Eprint [0]{\href }%
\providecommand \doibase [0]{http://dx.doi.org/}%
\providecommand \selectlanguage [0]{\@gobble}%
\providecommand \bibinfo  [0]{\@secondoftwo}%
\providecommand \bibfield  [0]{\@secondoftwo}%
\providecommand \translation [1]{[#1]}%
\providecommand \BibitemOpen [0]{}%
\providecommand \bibitemStop [0]{}%
\providecommand \bibitemNoStop [0]{.\EOS\space}%
\providecommand \EOS [0]{\spacefactor3000\relax}%
\providecommand \BibitemShut  [1]{\csname bibitem#1\endcsname}%
\let\auto@bib@innerbib\@empty
\bibitem [{\citenamefont {O'Hern}\ \emph {et~al.}(2003)\citenamefont {O'Hern},
  \citenamefont {Silbert}, \citenamefont {Liu},\ and\ \citenamefont
  {Nagel}}]{Ohern2003}%
  \BibitemOpen
  \bibfield  {author} {\bibinfo {author} {\bibfnamefont {C.~S.}\ \bibnamefont
  {O'Hern}}, \bibinfo {author} {\bibfnamefont {L.~E.}\ \bibnamefont {Silbert}},
  \bibinfo {author} {\bibfnamefont {A.~J.}\ \bibnamefont {Liu}}, \ and\
  \bibinfo {author} {\bibfnamefont {S.~R.}\ \bibnamefont {Nagel}},\ }\href@noop
  {} {\bibfield  {journal} {\bibinfo  {journal} {Phys. Rev. E}\ }\textbf
  {\bibinfo {volume} {68}},\ \bibinfo {pages} {011306} (\bibinfo {year}
  {2003})}\BibitemShut {NoStop}%
\bibitem [{\citenamefont {Wyart}\ \emph
  {et~al.}(2005{\natexlab{a}})\citenamefont {Wyart}, \citenamefont {Nagel},\
  and\ \citenamefont {Witten}}]{Wyart2005}%
  \BibitemOpen
  \bibfield  {author} {\bibinfo {author} {\bibfnamefont {M.}~\bibnamefont
  {Wyart}}, \bibinfo {author} {\bibfnamefont {S.~R.}\ \bibnamefont {Nagel}}, \
  and\ \bibinfo {author} {\bibfnamefont {T.~A.}\ \bibnamefont {Witten}},\
  }\href@noop {} {\bibfield  {journal} {\bibinfo  {journal} {Europhys. Lett.}\
  }\textbf {\bibinfo {volume} {72}},\ \bibinfo {pages} {486} (\bibinfo {year}
  {2005}{\natexlab{a}})}\BibitemShut {NoStop}%
\bibitem [{\citenamefont {Goodrich}\ \emph {et~al.}(2013)\citenamefont
  {Goodrich}, \citenamefont {Ellenbroek},\ and\ \citenamefont
  {Liu}}]{Goodrich2013}%
  \BibitemOpen
  \bibfield  {author} {\bibinfo {author} {\bibfnamefont {C.~P.}\ \bibnamefont
  {Goodrich}}, \bibinfo {author} {\bibfnamefont {W.~G.}\ \bibnamefont
  {Ellenbroek}}, \ and\ \bibinfo {author} {\bibfnamefont {A.~J.}\ \bibnamefont
  {Liu}},\ }\href@noop {} {\bibfield  {journal} {\bibinfo  {journal} {Soft
  Matter}\ }\textbf {\bibinfo {volume} {9}},\ \bibinfo {pages} {10993}
  (\bibinfo {year} {2013})}\BibitemShut {NoStop}%
\bibitem [{\citenamefont {Schoenholz}\ \emph {et~al.}(2013)\citenamefont
  {Schoenholz}, \citenamefont {Goodrich}, \citenamefont {Kogan}, \citenamefont
  {Liu},\ and\ \citenamefont {Nagel}}]{Schoenholz2013}%
  \BibitemOpen
  \bibfield  {author} {\bibinfo {author} {\bibfnamefont {S.~S.}\ \bibnamefont
  {Schoenholz}}, \bibinfo {author} {\bibfnamefont {C.~P.}\ \bibnamefont
  {Goodrich}}, \bibinfo {author} {\bibfnamefont {O.}~\bibnamefont {Kogan}},
  \bibinfo {author} {\bibfnamefont {A.~J.}\ \bibnamefont {Liu}}, \ and\
  \bibinfo {author} {\bibfnamefont {S.~R.}\ \bibnamefont {Nagel}},\ }\href@noop
  {} {\bibfield  {journal} {\bibinfo  {journal} {Soft Matter}\ }\textbf
  {\bibinfo {volume} {9}},\ \bibinfo {pages} {11000} (\bibinfo {year}
  {2013})}\BibitemShut {NoStop}%
\bibitem [{\citenamefont {Sussman}\ \emph {et~al.}(2015)\citenamefont
  {Sussman}, \citenamefont {Goodrich}, \citenamefont {Liu},\ and\ \citenamefont
  {Nagel}}]{Sussman2015}%
  \BibitemOpen
  \bibfield  {author} {\bibinfo {author} {\bibfnamefont {D.~M.}\ \bibnamefont
  {Sussman}}, \bibinfo {author} {\bibfnamefont {C.~P.}\ \bibnamefont
  {Goodrich}}, \bibinfo {author} {\bibfnamefont {A.~J.}\ \bibnamefont {Liu}}, \
  and\ \bibinfo {author} {\bibfnamefont {S.~R.}\ \bibnamefont {Nagel}},\
  }\href@noop {} {\bibfield  {journal} {\bibinfo  {journal} {Soft Matter}\
  }\textbf {\bibinfo {volume} {11}},\ \bibinfo {pages} {2745} (\bibinfo {year}
  {2015})}\BibitemShut {NoStop}%
\bibitem [{\citenamefont {Ellenbroek}\ \emph {et~al.}(2006)\citenamefont
  {Ellenbroek}, \citenamefont {Somfai}, \citenamefont {van Hecke},\ and\
  \citenamefont {van Saarloos}}]{Ellenbroek2006}%
  \BibitemOpen
  \bibfield  {author} {\bibinfo {author} {\bibfnamefont {W.~G.}\ \bibnamefont
  {Ellenbroek}}, \bibinfo {author} {\bibfnamefont {E.}~\bibnamefont {Somfai}},
  \bibinfo {author} {\bibfnamefont {M.}~\bibnamefont {van Hecke}}, \ and\
  \bibinfo {author} {\bibfnamefont {W.}~\bibnamefont {van Saarloos}},\
  }\href@noop {} {\bibfield  {journal} {\bibinfo  {journal} {Phys. Rev. Lett.}\
  }\textbf {\bibinfo {volume} {97}},\ \bibinfo {pages} {258001} (\bibinfo
  {year} {2006})}\BibitemShut {NoStop}%
\bibitem [{\citenamefont {M.Wyart}(2010)}]{Wyart2010}%
  \BibitemOpen
  \bibfield  {author} {\bibinfo {author} {\bibnamefont {M.Wyart}},\ }\href@noop
  {} {\bibfield  {journal} {\bibinfo  {journal} {Europhys. Lett.}\ }\textbf
  {\bibinfo {volume} {89}},\ \bibinfo {pages} {64001} (\bibinfo {year}
  {2010})}\BibitemShut {NoStop}%
\bibitem [{\citenamefont {Lerner}\ \emph {et~al.}(2014)\citenamefont {Lerner},
  \citenamefont {DeGiulu}, \citenamefont {D{\"u}ring},\ and\ \citenamefont
  {Wyart}}]{Lerner2014}%
  \BibitemOpen
  \bibfield  {author} {\bibinfo {author} {\bibfnamefont {E.}~\bibnamefont
  {Lerner}}, \bibinfo {author} {\bibfnamefont {E.}~\bibnamefont {DeGiulu}},
  \bibinfo {author} {\bibfnamefont {G.}~\bibnamefont {D{\"u}ring}}, \ and\
  \bibinfo {author} {\bibfnamefont {M.}~\bibnamefont {Wyart}},\ }\href@noop {}
  {\bibfield  {journal} {\bibinfo  {journal} {Soft Matter}\ }\textbf {\bibinfo
  {volume} {10}},\ \bibinfo {pages} {5085} (\bibinfo {year}
  {2014})}\BibitemShut {NoStop}%
\bibitem [{\citenamefont {Upadhyaya}\ \emph {et~al.}(2014)\citenamefont
  {Upadhyaya}, \citenamefont {G\'omez},\ and\ \citenamefont
  {Vitelli}}]{Upadhyaya2014}%
  \BibitemOpen
  \bibfield  {author} {\bibinfo {author} {\bibfnamefont {N.}~\bibnamefont
  {Upadhyaya}}, \bibinfo {author} {\bibfnamefont {L.~R.}\ \bibnamefont
  {G\'omez}}, \ and\ \bibinfo {author} {\bibfnamefont {V.}~\bibnamefont
  {Vitelli}},\ }\href@noop {} {\bibfield  {journal} {\bibinfo  {journal} {Phys.
  Rev. X}\ }\textbf {\bibinfo {volume} {4}},\ \bibinfo {pages} {011045}
  (\bibinfo {year} {2014})}\BibitemShut {NoStop}%
\bibitem [{\citenamefont {Goodrich}\ \emph {et~al.}(2015)\citenamefont
  {Goodrich}, \citenamefont {Liu},\ and\ \citenamefont {Nagel}}]{Goodrich2015}%
  \BibitemOpen
  \bibfield  {author} {\bibinfo {author} {\bibfnamefont {C.~P.}\ \bibnamefont
  {Goodrich}}, \bibinfo {author} {\bibfnamefont {A.~J.}\ \bibnamefont {Liu}}, \
  and\ \bibinfo {author} {\bibfnamefont {S.~R.}\ \bibnamefont {Nagel}},\
  }\href@noop {} {\bibfield  {journal} {\bibinfo  {journal} {Phys. Rev. Lett.}\
  }\textbf {\bibinfo {volume} {114}},\ \bibinfo {pages} {225501} (\bibinfo
  {year} {2015})}\BibitemShut {NoStop}%
\bibitem [{\citenamefont {Paulose}\ \emph {et~al.}(2015)\citenamefont
  {Paulose}, \citenamefont {Meeussen},\ and\ \citenamefont
  {Vitelli}}]{Paulose2015}%
  \BibitemOpen
  \bibfield  {author} {\bibinfo {author} {\bibfnamefont {J.}~\bibnamefont
  {Paulose}}, \bibinfo {author} {\bibfnamefont {A.~S.}\ \bibnamefont
  {Meeussen}}, \ and\ \bibinfo {author} {\bibfnamefont {V.}~\bibnamefont
  {Vitelli}},\ }\href@noop {} {\bibfield  {journal} {\bibinfo  {journal} {Proc.
  Natl. Acad. Sci. USA}\ }\textbf {\bibinfo {volume} {112}},\ \bibinfo {pages}
  {7639} (\bibinfo {year} {2015})}\BibitemShut {NoStop}%
\bibitem [{\citenamefont {Florijn}\ \emph {et~al.}(2014)\citenamefont
  {Florijn}, \citenamefont {Coulais},\ and\ \citenamefont {van
  Hecke}}]{Florijn2014}%
  \BibitemOpen
  \bibfield  {author} {\bibinfo {author} {\bibfnamefont {B.}~\bibnamefont
  {Florijn}}, \bibinfo {author} {\bibfnamefont {C.}~\bibnamefont {Coulais}}, \
  and\ \bibinfo {author} {\bibfnamefont {M.}~\bibnamefont {van Hecke}},\
  }\href@noop {} {\bibfield  {journal} {\bibinfo  {journal} {Phys. Rev. Lett.}\
  }\textbf {\bibinfo {volume} {113}},\ \bibinfo {pages} {175503} (\bibinfo
  {year} {2014})}\BibitemShut {NoStop}%
\bibitem [{\citenamefont {Wyart}(2005)}]{Wyart:2005vu}%
  \BibitemOpen
  \bibfield  {author} {\bibinfo {author} {\bibfnamefont {M.}~\bibnamefont
  {Wyart}},\ }\href@noop {} {\bibfield  {journal} {\bibinfo  {journal} {Ann
  Phys-Paris}\ }\textbf {\bibinfo {volume} {30}},\ \bibinfo {pages} {1}
  (\bibinfo {year} {2005})}\BibitemShut {NoStop}%
\bibitem [{\citenamefont {Ostojic}\ \emph {et~al.}(2006)\citenamefont
  {Ostojic}, \citenamefont {Somfai},\ and\ \citenamefont
  {Nienhuis}}]{Nienhuis2006}%
  \BibitemOpen
  \bibfield  {author} {\bibinfo {author} {\bibfnamefont {S.}~\bibnamefont
  {Ostojic}}, \bibinfo {author} {\bibfnamefont {E.}~\bibnamefont {Somfai}}, \
  and\ \bibinfo {author} {\bibfnamefont {B.}~\bibnamefont {Nienhuis}},\
  }\href@noop {} {\bibfield  {journal} {\bibinfo  {journal} {Nature}\ }\textbf
  {\bibinfo {volume} {439}},\ \bibinfo {pages} {828} (\bibinfo {year}
  {2006})}\BibitemShut {NoStop}%
\bibitem [{\citenamefont {Kovalcinova}\ \emph {et~al.}()\citenamefont
  {Kovalcinova}, \citenamefont {Goullet},\ and\ \citenamefont
  {Kondic}}]{Kondic2015}%
  \BibitemOpen
  \bibfield  {author} {\bibinfo {author} {\bibfnamefont {L.}~\bibnamefont
  {Kovalcinova}}, \bibinfo {author} {\bibfnamefont {A.}~\bibnamefont
  {Goullet}}, \ and\ \bibinfo {author} {\bibfnamefont {L.}~\bibnamefont
  {Kondic}},\ }\href@noop {} {\bibinfo  {journal}
  {http://arxiv.org/abs/1511.05556}\ }\BibitemShut {NoStop}%
\bibitem [{\citenamefont {Calladine}(1978)}]{Calladine1978}%
  \BibitemOpen
\bibfield  {journal} {  }\bibfield  {author} {\bibinfo {author} {\bibfnamefont
  {C.~R.}\ \bibnamefont {Calladine}},\ }\href@noop {} {\bibfield  {journal}
  {\bibinfo  {journal} {Int. J. Solids and Structures}\ }\textbf {\bibinfo
  {volume} {14}},\ \bibinfo {pages} {161} (\bibinfo {year} {1978})}\BibitemShut
  {NoStop}%
\bibitem [{\citenamefont {Maxwell}(1865)}]{Maxwell1865}%
  \BibitemOpen
  \bibfield  {author} {\bibinfo {author} {\bibfnamefont {J.~C.}\ \bibnamefont
  {Maxwell}},\ }\href@noop {} {\bibfield  {journal} {\bibinfo  {journal}
  {Philosophical Magazine}\ }\textbf {\bibinfo {volume} {27}},\ \bibinfo
  {pages} {294} (\bibinfo {year} {1865})}\BibitemShut {NoStop}%
\bibitem [{\citenamefont {Alexander}(1998)}]{Alexander1998}%
  \BibitemOpen
  \bibfield  {author} {\bibinfo {author} {\bibfnamefont {S.}~\bibnamefont
  {Alexander}},\ }\href@noop {} {\bibfield  {journal} {\bibinfo  {journal}
  {Physics Reports}\ }\textbf {\bibinfo {volume} {296}},\ \bibinfo {pages} {65}
  (\bibinfo {year} {1998})}\BibitemShut {NoStop}%
\bibitem [{\citenamefont {Wyart}\ \emph
  {et~al.}(2005{\natexlab{b}})\citenamefont {Wyart}, \citenamefont {Silbert},
  \citenamefont {Nagel},\ and\ \citenamefont {Witten}}]{Silbert2005pre}%
  \BibitemOpen
  \bibfield  {author} {\bibinfo {author} {\bibfnamefont {M.}~\bibnamefont
  {Wyart}}, \bibinfo {author} {\bibfnamefont {L.~E.}\ \bibnamefont {Silbert}},
  \bibinfo {author} {\bibfnamefont {S.~R.}\ \bibnamefont {Nagel}}, \ and\
  \bibinfo {author} {\bibfnamefont {T.~A.}\ \bibnamefont {Witten}},\
  }\href@noop {} {\bibfield  {journal} {\bibinfo  {journal} {Phys. Rev. E}\
  }\textbf {\bibinfo {volume} {72}},\ \bibinfo {pages} {051306} (\bibinfo
  {year} {2005}{\natexlab{b}})}\BibitemShut {NoStop}%
\bibitem [{\citenamefont {Pellegrino}(1993)}]{Pellegrino1993}%
  \BibitemOpen
  \bibfield  {author} {\bibinfo {author} {\bibfnamefont {S.}~\bibnamefont
  {Pellegrino}},\ }\href@noop {} {\bibfield  {journal} {\bibinfo  {journal}
  {Int. J. Solids Struct.}\ }\textbf {\bibinfo {volume} {30}},\ \bibinfo
  {pages} {3025} (\bibinfo {year} {1993})}\BibitemShut {NoStop}%
\bibitem [{\citenamefont {Lerner}\ \emph {et~al.}(2012)\citenamefont {Lerner},
  \citenamefont {D{\"u}ring},\ and\ \citenamefont {Wyart}}]{Wyart2012}%
  \BibitemOpen
  \bibfield  {author} {\bibinfo {author} {\bibfnamefont {E.}~\bibnamefont
  {Lerner}}, \bibinfo {author} {\bibfnamefont {G.}~\bibnamefont {D{\"u}ring}},
  \ and\ \bibinfo {author} {\bibfnamefont {M.}~\bibnamefont {Wyart}},\
  }\href@noop {} {\bibfield  {journal} {\bibinfo  {journal} {Proc. Natl. Acad.
  Sci. USA}\ }\textbf {\bibinfo {volume} {109}},\ \bibinfo {pages} {4798}
  (\bibinfo {year} {2012})}\BibitemShut {NoStop}%
\bibitem [{\citenamefont {Lubensky}\ \emph {et~al.}(2015)\citenamefont
  {Lubensky}, \citenamefont {Kane}, \citenamefont {Mao}, \citenamefont
  {Souslov},\ and\ \citenamefont {Sun}}]{LubenskyRev}%
  \BibitemOpen
  \bibfield  {author} {\bibinfo {author} {\bibfnamefont {T.~C.}\ \bibnamefont
  {Lubensky}}, \bibinfo {author} {\bibfnamefont {C.~L.}\ \bibnamefont {Kane}},
  \bibinfo {author} {\bibfnamefont {X.}~\bibnamefont {Mao}}, \bibinfo {author}
  {\bibfnamefont {A.}~\bibnamefont {Souslov}}, \ and\ \bibinfo {author}
  {\bibfnamefont {K.}~\bibnamefont {Sun}},\ }\href@noop {} {\bibfield
  {journal} {\bibinfo  {journal} {Rep. Prog. Phys.}\ }\textbf {\bibinfo
  {volume} {78}},\ \bibinfo {pages} {073901} (\bibinfo {year}
  {2015})}\BibitemShut {NoStop}%
\bibitem [{\citenamefont {Kane}\ and\ \citenamefont
  {Lubensky}(2014)}]{Lubensky2014}%
  \BibitemOpen
  \bibfield  {author} {\bibinfo {author} {\bibfnamefont {C.~L.}\ \bibnamefont
  {Kane}}\ and\ \bibinfo {author} {\bibfnamefont {T.~C.}\ \bibnamefont
  {Lubensky}},\ }\href@noop {} {\bibfield  {journal} {\bibinfo  {journal}
  {Nature Physics}\ }\textbf {\bibinfo {volume} {10}},\ \bibinfo {pages} {39}
  (\bibinfo {year} {2014})}\BibitemShut {NoStop}%
\bibitem [{\citenamefont {Sussman}\ \emph {et~al.}()\citenamefont {Sussman},
  \citenamefont {Stenull},\ and\ \citenamefont {Lubensky}}]{SussmanTopo}%
  \BibitemOpen
  \bibfield  {author} {\bibinfo {author} {\bibfnamefont {D.~M.}\ \bibnamefont
  {Sussman}}, \bibinfo {author} {\bibfnamefont {O.}~\bibnamefont {Stenull}}, \
  and\ \bibinfo {author} {\bibfnamefont {T.~C.}\ \bibnamefont {Lubensky}},\
  }\href@noop {} {\bibinfo  {journal} {http://arxiv.org/abs/1512.04480}\
  }\BibitemShut {NoStop}%
\bibitem [{\citenamefont {Driscoll}\ \emph {et~al.}(2015)\citenamefont
  {Driscoll}, \citenamefont {Chen}, \citenamefont {Beuman}, \citenamefont
  {Ulrich}, \citenamefont {Nagel},\ and\ \citenamefont
  {Vitelli}}]{Driscoll:2015vf}%
  \BibitemOpen
\bibfield  {journal} {  }\bibfield  {author} {\bibinfo {author} {\bibfnamefont
  {M.~M.}\ \bibnamefont {Driscoll}}, \bibinfo {author} {\bibfnamefont
  {B.~G.-G.}\ \bibnamefont {Chen}}, \bibinfo {author} {\bibfnamefont {T.~H.}\
  \bibnamefont {Beuman}}, \bibinfo {author} {\bibfnamefont {S.}~\bibnamefont
  {Ulrich}}, \bibinfo {author} {\bibfnamefont {S.~R.}\ \bibnamefont {Nagel}}, \
  and\ \bibinfo {author} {\bibfnamefont {V.}~\bibnamefont {Vitelli}},\
  }\href@noop {} {\bibfield  {journal} {\bibinfo  {journal} {arXiv}\ }
  (\bibinfo {year} {2015})},\ \Eprint {http://arxiv.org/abs/1501.04227v1}
  {1501.04227v1} \BibitemShut {NoStop}%
\bibitem [{\citenamefont {Yan}\ and\ \citenamefont {Wyart}(2016)}]{Wyart2016}%
  \BibitemOpen
  \bibfield  {author} {\bibinfo {author} {\bibfnamefont {L.}~\bibnamefont
  {Yan}}\ and\ \bibinfo {author} {\bibfnamefont {M.}~\bibnamefont {Wyart}},\
  }\href@noop {} {\bibfield  {journal} {\bibinfo  {journal} {arXiv}\ }
  (\bibinfo {year} {2016})},\ \Eprint {http://arxiv.org/abs/1601.02141}
  {1601.02141} \BibitemShut {NoStop}%
\bibitem [{\citenamefont {Bitzec}\ \emph {et~al.}(2006)\citenamefont {Bitzec},
  \citenamefont {Koshkinen}, \citenamefont {G{\"a}hler}, \citenamefont
  {Moseler},\ and\ \citenamefont {Gumbsch}}]{quench}%
  \BibitemOpen
  \bibfield  {author} {\bibinfo {author} {\bibfnamefont {E.}~\bibnamefont
  {Bitzec}}, \bibinfo {author} {\bibfnamefont {P.}~\bibnamefont {Koshkinen}},
  \bibinfo {author} {\bibfnamefont {F.}~\bibnamefont {G{\"a}hler}}, \bibinfo
  {author} {\bibfnamefont {M.}~\bibnamefont {Moseler}}, \ and\ \bibinfo
  {author} {\bibfnamefont {P.}~\bibnamefont {Gumbsch}},\ }\href@noop {}
  {\bibfield  {journal} {\bibinfo  {journal} {Phys. Rev. Lett.}\ }\textbf
  {\bibinfo {volume} {97}},\ \bibinfo {pages} {170201} (\bibinfo {year}
  {2006})}\BibitemShut {NoStop}%
\bibitem [{\citenamefont {Lehoucq}\ \emph {et~al.}(1998)\citenamefont
  {Lehoucq}, \citenamefont {Sorensen},\ and\ \citenamefont {Yang}}]{arpack}%
  \BibitemOpen
  \bibfield  {author} {\bibinfo {author} {\bibfnamefont {R.}~\bibnamefont
  {Lehoucq}}, \bibinfo {author} {\bibfnamefont {D.~C.}\ \bibnamefont
  {Sorensen}}, \ and\ \bibinfo {author} {\bibfnamefont {C.}~\bibnamefont
  {Yang}},\ }\href@noop {} {\emph {\bibinfo {title} {ARPACK User's Guide:
  Solution of Large-Scale Eigenvalue Problems With Implicitly Restorted Arnoldi
  Methods}}}\ (\bibinfo  {publisher} {SIAM},\ \bibinfo {address}
  {Philadelphia},\ \bibinfo {year} {1998})\BibitemShut {NoStop}%
\bibitem [{\citenamefont {D{\"u}ring}\ \emph {et~al.}(2013)\citenamefont
  {D{\"u}ring}, \citenamefont {Lerner},\ and\ \citenamefont
  {Wyart}}]{During2013}%
  \BibitemOpen
  \bibfield  {author} {\bibinfo {author} {\bibfnamefont {G.}~\bibnamefont
  {D{\"u}ring}}, \bibinfo {author} {\bibfnamefont {E.}~\bibnamefont {Lerner}},
  \ and\ \bibinfo {author} {\bibfnamefont {M.}~\bibnamefont {Wyart}},\
  }\href@noop {} {\bibfield  {journal} {\bibinfo  {journal} {Soft Matter}\
  }\textbf {\bibinfo {volume} {9}},\ \bibinfo {pages} {146} (\bibinfo {year}
  {2013})}\BibitemShut {NoStop}%
\bibitem [{\citenamefont {Silbert}\ \emph {et~al.}(2005)\citenamefont
  {Silbert}, \citenamefont {Liu},\ and\ \citenamefont {Nagel}}]{Silbert2005}%
  \BibitemOpen
  \bibfield  {author} {\bibinfo {author} {\bibfnamefont {L.~E.}\ \bibnamefont
  {Silbert}}, \bibinfo {author} {\bibfnamefont {A.~J.}\ \bibnamefont {Liu}}, \
  and\ \bibinfo {author} {\bibfnamefont {S.~R.}\ \bibnamefont {Nagel}},\
  }\href@noop {} {\bibfield  {journal} {\bibinfo  {journal} {Phys. Rev. Lett.}\
  }\textbf {\bibinfo {volume} {95}},\ \bibinfo {pages} {098301} (\bibinfo
  {year} {2005})}\BibitemShut {NoStop}%
\bibitem [{\citenamefont {Tighe}\ \emph {et~al.}(2008)\citenamefont {Tighe},
  \citenamefont {van Eerd},\ and\ \citenamefont {Vlugt}}]{Tighe2008}%
  \BibitemOpen
  \bibfield  {author} {\bibinfo {author} {\bibfnamefont {B.~P.}\ \bibnamefont
  {Tighe}}, \bibinfo {author} {\bibfnamefont {A.~R.~T.}\ \bibnamefont {van
  Eerd}}, \ and\ \bibinfo {author} {\bibfnamefont {T.~J.~H.}\ \bibnamefont
  {Vlugt}},\ }\href@noop {} {\bibfield  {journal} {\bibinfo  {journal} {Phys.
  Rev. Lett.}\ }\textbf {\bibinfo {volume} {100}},\ \bibinfo {pages} {238001}
  (\bibinfo {year} {2008})}\BibitemShut {NoStop}%
\bibitem [{\citenamefont {Tighe}\ \emph {et~al.}(2010)\citenamefont {Tighe},
  \citenamefont {Snoeijer}, \citenamefont {Vlugt},\ and\ \citenamefont {van
  Hecke}}]{Tighe2010}%
  \BibitemOpen
  \bibfield  {author} {\bibinfo {author} {\bibfnamefont {B.~P.}\ \bibnamefont
  {Tighe}}, \bibinfo {author} {\bibfnamefont {J.~H.}\ \bibnamefont {Snoeijer}},
  \bibinfo {author} {\bibfnamefont {T.~J.~H.}\ \bibnamefont {Vlugt}}, \ and\
  \bibinfo {author} {\bibfnamefont {M.}~\bibnamefont {van Hecke}},\ }\href@noop
  {} {\bibfield  {journal} {\bibinfo  {journal} {Soft Matter}\ }\textbf
  {\bibinfo {volume} {6}},\ \bibinfo {pages} {2908} (\bibinfo {year}
  {2010})}\BibitemShut {NoStop}%
\bibitem [{\citenamefont {Tighe}\ and\ \citenamefont
  {Vlugt}(2011)}]{Tighe2011}%
  \BibitemOpen
  \bibfield  {author} {\bibinfo {author} {\bibfnamefont {B.~P.}\ \bibnamefont
  {Tighe}}\ and\ \bibinfo {author} {\bibfnamefont {T.~J.~H.}\ \bibnamefont
  {Vlugt}},\ }\href@noop {} {\bibfield  {journal} {\bibinfo  {journal} {J.
  Stat. Mech.: Theory and Exp.}\ }\textbf {\bibinfo {volume} {11}},\ \bibinfo
  {pages} {1742} (\bibinfo {year} {2011})}\BibitemShut {NoStop}%
\bibitem [{\citenamefont {Kondic}\ \emph {et~al.}(2012)\citenamefont {Kondic},
  \citenamefont {Goullet}, \citenamefont {Kramar}, \citenamefont {Mischaikow},\
  and\ \citenamefont {Behringer}}]{perhom}%
  \BibitemOpen
  \bibfield  {author} {\bibinfo {author} {\bibfnamefont {L.}~\bibnamefont
  {Kondic}}, \bibinfo {author} {\bibfnamefont {C.~S.}\ \bibnamefont {Goullet},
  \bibfnamefont {A.~amd~O'Hern}}, \bibinfo {author} {\bibfnamefont
  {M.}~\bibnamefont {Kramar}}, \bibinfo {author} {\bibfnamefont
  {K.}~\bibnamefont {Mischaikow}}, \ and\ \bibinfo {author} {\bibfnamefont
  {R.~P.}\ \bibnamefont {Behringer}},\ }\href@noop {} {\bibfield  {journal}
  {\bibinfo  {journal} {Europhys. Lett.}\ }\textbf {\bibinfo {volume} {97}},\
  \bibinfo {pages} {54001} (\bibinfo {year} {2012})}\BibitemShut {NoStop}%
\bibitem [{\citenamefont {Bassett}\ \emph {et~al.}(2012)\citenamefont
  {Bassett}, \citenamefont {Owens}, \citenamefont {Daniels},\ and\
  \citenamefont {Porter}}]{Bassett2012}%
  \BibitemOpen
  \bibfield  {author} {\bibinfo {author} {\bibfnamefont {D.~S.}\ \bibnamefont
  {Bassett}}, \bibinfo {author} {\bibfnamefont {E.~T.}\ \bibnamefont {Owens}},
  \bibinfo {author} {\bibfnamefont {K.~E.}\ \bibnamefont {Daniels}}, \ and\
  \bibinfo {author} {\bibfnamefont {M.~A.}\ \bibnamefont {Porter}},\
  }\href@noop {} {\bibfield  {journal} {\bibinfo  {journal} {Phys. Rev. E}\
  }\textbf {\bibinfo {volume} {86}},\ \bibinfo {pages} {041306} (\bibinfo
  {year} {2012})}\BibitemShut {NoStop}%
\bibitem [{\citenamefont {Katifori}\ and\ \citenamefont
  {Magnasco}(2012)}]{Katifori2012}%
  \BibitemOpen
  \bibfield  {author} {\bibinfo {author} {\bibfnamefont {E.}~\bibnamefont
  {Katifori}}\ and\ \bibinfo {author} {\bibfnamefont {M.~O.}\ \bibnamefont
  {Magnasco}},\ }\href@noop {} {\bibfield  {journal} {\bibinfo  {journal} {PLoS
  ONE}\ }\textbf {\bibinfo {volume} {7}},\ \bibinfo {pages} {e37994} (\bibinfo
  {year} {2012})}\BibitemShut {NoStop}%
\bibitem [{\citenamefont {Modes}\ \emph {et~al.}()\citenamefont {Modes},
  \citenamefont {Magnasco},\ and\ \citenamefont {Katifori}}]{Katifori2015}%
  \BibitemOpen
  \bibfield  {author} {\bibinfo {author} {\bibfnamefont {C.~D.}\ \bibnamefont
  {Modes}}, \bibinfo {author} {\bibfnamefont {M.~O.}\ \bibnamefont {Magnasco}},
  \ and\ \bibinfo {author} {\bibfnamefont {E.}~\bibnamefont {Katifori}},\
  }\href@noop {} {\bibinfo  {journal} {http://arxiv.org/abs/1410.3951}\
  }\BibitemShut {NoStop}%
\end{thebibliography}%

\appendix*

\section{Participation-ratio-minimized SSS}
Although the SSS $e$ described in the main text is the unique element that would vanish from the null space if bond $b_i$ were removed, it is not the only SSS that involves the bond $b_i$. In particular, any linear combination of $e$ and the other elements of $\textrm{ker}(\mathcal{N})$ are equally valid and may be involved in the real response of, e.g., spring networks to an imposed stress on bond $b_i$. Since different physical measurements made on these SSS may lead to different responses, it is important to ask if the $e$ generated by the above procedure are generically representative of bond-localized states of self stress. For instance, one could ask if the states of self stress generated by the above procedure are maximally localized, or if some linear combination of the states of self stress yields a more tightly localized set of stresses. One approach, alluded to in the main text, would be to modify the method of D{\"u}ring et al. \cite{During2013} and study dashpot systems perturbed by an instantaneous strain of a bond.

Here we employ an alternate approach, directly studying whether linear combinations of modes can lead to a more localized SSS centered on bond $b_i$. We first identify the participation ratio of $e$ defined in the text, and then attempt to find a nearby local minimum of $p_r$ in the space of linear combinations of the elements of $\textrm{ker}(\mathcal{N})$ using the FIRE algorithm \cite{quench}. We find that we are typically able to find linear combinations of SSS, which we denote $\tilde{e}$, that are as localized as the lowest-participation-ratio portion of the distribution of $e$. Since this minimization scheme can take the system out of local minima (i.e., to nearby minima in which the dominant bond is no longer the bond selected to construct $e$), we additionally require that $\tilde{e}$ remain close to the original state $e$ by setting a lower bound on the acceptable overlap between the two states, $e\cdot \tilde{e}$, for $\tilde{e}$ to be included in the averages over states described below. 

\begin{figure}
\centerline{\includegraphics[width=.8\linewidth]{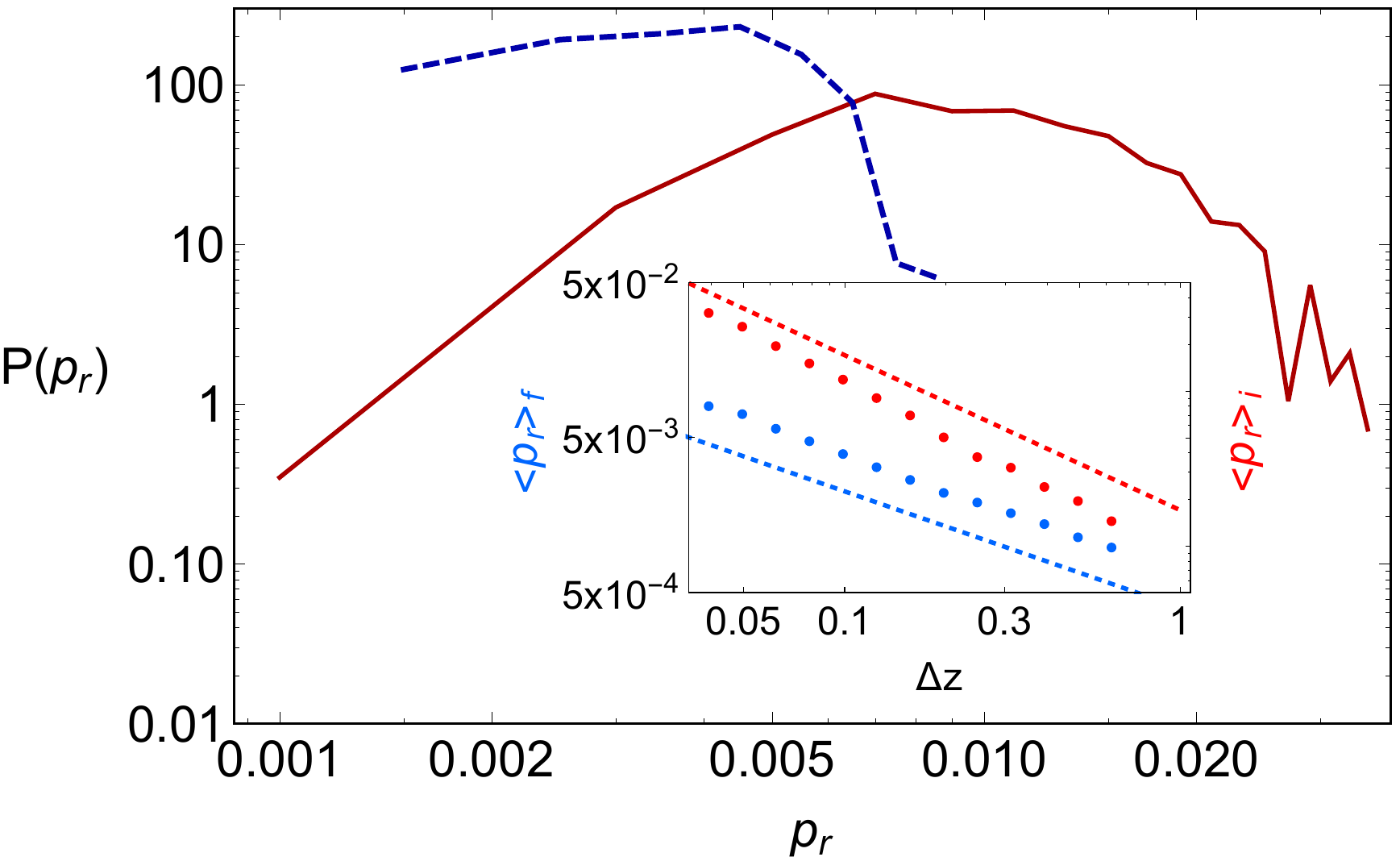}}
\caption{\label{fig:prplotsupmat} Distribution of participation ratios for the states of self stress $e$ (right, solid curve) and $\tilde{e}$ (left, dashed curve), for an $N=8192$ system in two dimensions prepared at $\Delta z=0.099$. Inset. The mean $P(p_r)$ averaged over the $e$ (upper, red points) and $\tilde{e}$ (lower, blue points) as a function of $\Delta z$. The dotted lines are guides to the eye with slope $-1$ (upper) and $-3/4$ (lower).}
\end{figure}

\begin{figure}
\centerline{\includegraphics[width=.8\linewidth]{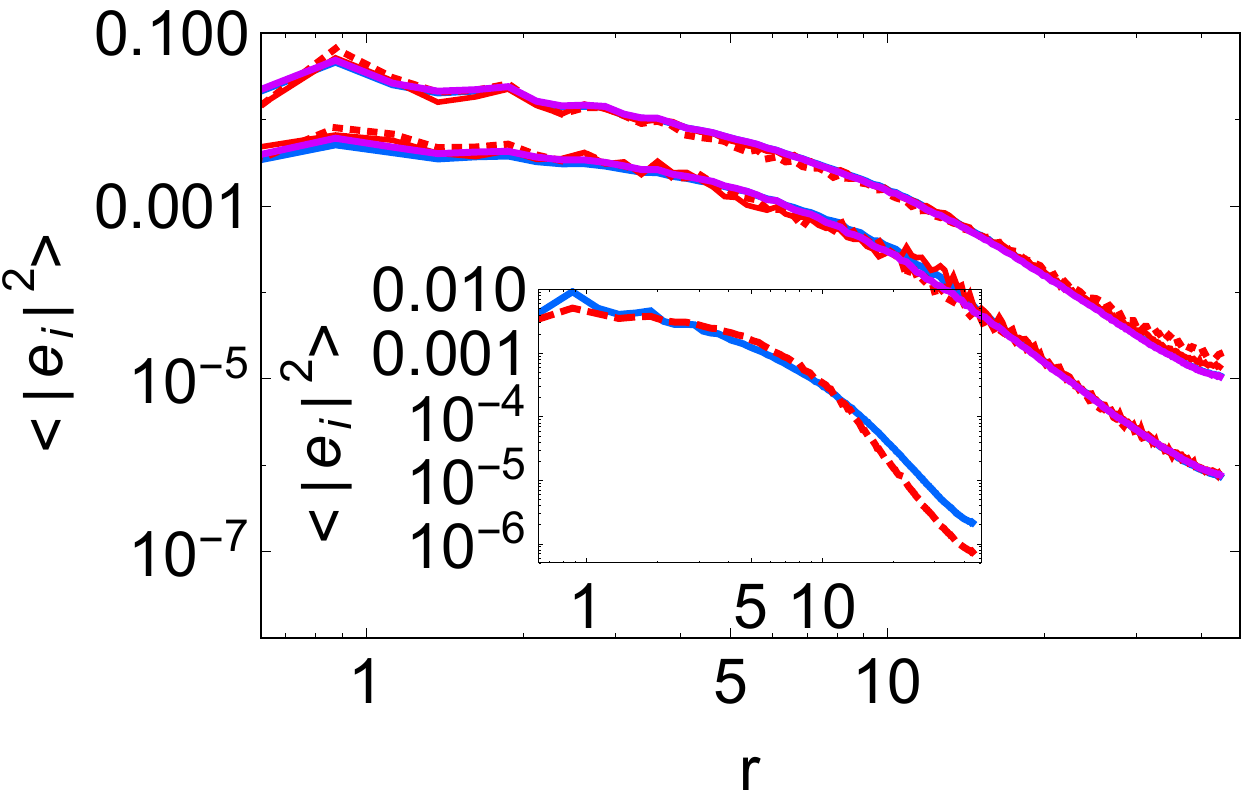}}
\caption{\label{fig:explot2dcombsupmat} Average radial profiles of states of self stress unique to bonds in a 2D system with $N=8192$ prepared at $\Delta z =0.31$. Blue lines correspond to averages over all bonds in the system, purple lines correspond to averages over those bonds whose state of self stress has a low participation ratio (lower 80th percentile), solid red lines correspond to averages over bonds with a large contribution to the bulk modulus (upper 5th percentile), and dotted red lines correspond to averages over bonds with a large contribution to the shear modulus (upper 5th percentile). The upper set of curves (vertically shifted by a factor of 5 for clarity) correspond to the averages over $e$, the lower set of curves are averages over $\tilde{e}$. Inset. Direct comparison between $e$ (blue, solid curve) and $\tilde{e}$ (red, dashed curve) averaged over all bonds in the system.}
\end{figure}

A representative plot demonstrating the change in participation ratio achieved by this method is shown in Fig. \ref{fig:prplotsupmat}. Of particular note is that the process of performing this participation-ratio-minimizing step changes the scaling of the mean and (not shown) standard deviation of the distribution of $p_r$ as a function of pressure. In accord with the exponents observed in the scaling collapse in the main text, we find that, e.g., $\langle p_r \rangle \sim \Delta z^{-\alpha}$ for $1/2 < \alpha < 1$ when the average is carried out over the $\tilde{e}$. In Fig. \ref{fig:explot2dcombsupmat} we plots the stress profiles of $e$ and $\tilde{e}$ as a function of distance from the selected bond averaged over different subsets of bonds in the system for a particular pressure. The figure shows the modestly tighter localization we generically find when studying $\tilde{e}$ as opposed to $e$. We also note that performing the local minimization of $p_r$ to obtain $\tilde{e}$ suppresses the distinction between bonds that contribute differently to the modulus. In effect, all $p_r$-minimized SSS, $\tilde{e}$, are approximately as localized as the most localized SSS $e$.

\begin{figure}
\centerline{\includegraphics[width=.5\linewidth]{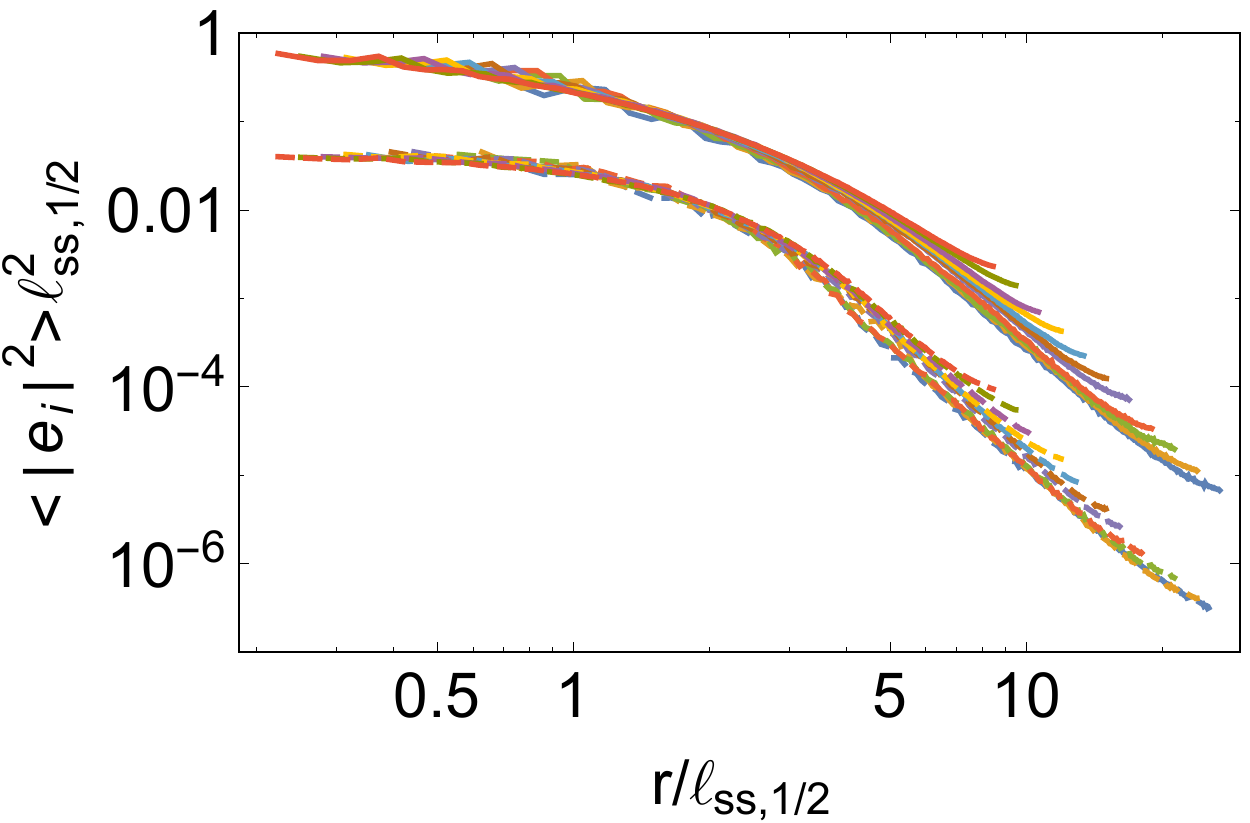}\includegraphics[width=.5\linewidth]{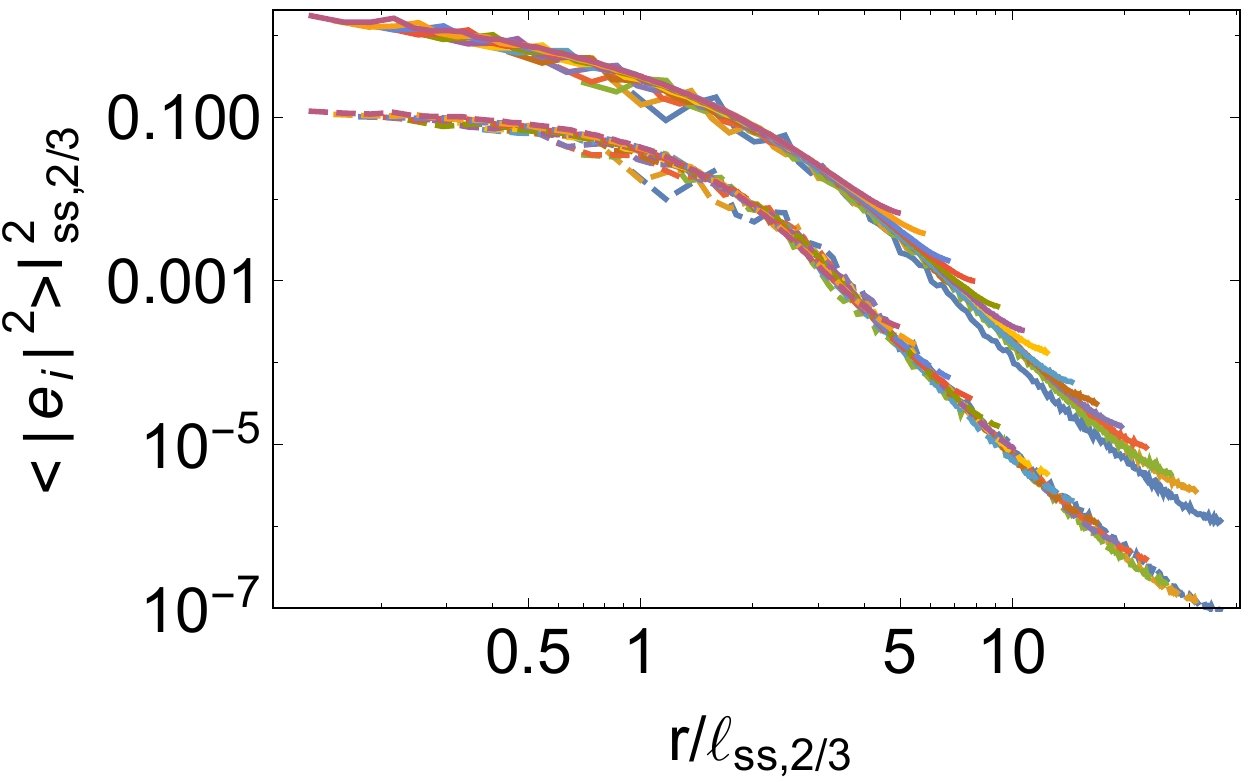}}
\caption{\label{fig:collapse2dsupmat} (Left) Average radial stress profiles for 2D systems prepared at $\Delta z =0.039 - 0.99$ as a function of $r/\ell_{ss,\alpha}$ for $\alpha=1/2$. Upper curves (vertically shifted by a factor of 10 for clarity) correspond to the states of self stress obtained by orthonormalizing the null space, $e$, and the lower set of curves are the same data after the participation-ratio-minimizing scheme has been applied, $\tilde{e}$. (Right) The same data presented as a function of $r/\ell_{ss,\alpha=2/3}$. }
\end{figure}

We note that unlike the $e$, the functional form of the average spatial profile of the $\tilde{e}$ is not well-described by a stretched exponential decay. For 3D systems the cost of performing a participation-minimizing step to find $\tilde{e}$ is prohibitive, but we have confirmed with $N=4096$ systems in $d=3$ that the qualitative differences between $e$ and $\tilde{e}$ are similar to what we found in $d=2$.  Even though the functional form of the spatial profile is different for $e$ and for $\tilde e$, however, the length scale $\ell_{ss}$ controlling scaling collapse of the spatial profile remains robust.  To demonstrate this robustness, in Fig. \ref{fig:collapse2dsupmat} we show the average radial stress profiles of the states $e$ and $\tilde{e}$ scaled by $\ell_{ss,\alpha}$ for two choices of $\alpha$. We find that the $\tilde{e}$ are collapsed to a very similar degree as $e$ by the different choices of $\alpha$. Of note, even though the $\tilde{e}$ are more tightly localized, and have a modestly different functional form, we find that they are well-collapsed by the $\ell_{ss,\alpha}$ at the same value of $\alpha$, or equivalently, by the same power of $\Delta z$, as the states $e$ themselves.  Thus, the length scale $\ell_{ss}$ is robustly defined for bond-localized states of self stress for the two different ways in which we have extract them.

\end{document}